\documentclass[useAMS,usenatbib]{mn2e}

\usepackage{amssymb}
\usepackage{amsmath}
\usepackage{times}
\usepackage{graphicx}
\usepackage{wasysym}

\def\aap{A\&A}
\def\apj{ApJ}

\def\apjl{ApJ}
\def\pasa{PASA}
\def\mnras{MNRAS}
\def\araa{ARA\&A}
\def\aj{AJ}

\def\nat{Nat}

\def\apjs{ApJS}
\def\jcap{JCAP}        

\newcommand{\mincir}{\raise
-2.truept\hbox{\rlap{\hbox{$\sim$}}\raise5.truept 
\hbox{$<$}\ }}
\newcommand{\magcir}{\raise
-2.truept\hbox{\rlap{\hbox{$\sim$}}\raise5.truept
\hbox{$>$}\ }}
\newcommand{\minmag}{\raise-2.truept\hbox{\rlap{\hbox{$<$}}\raise
6.truept\hbox
{$>$}\ }}

\newcommand{\be}{\begin{equation}}
\newcommand{\ee}{\end{equation}}
\newcommand{\ba}{\begin{eqnarray}}
\newcommand{\ea}{\end{eqnarray}}
\newcommand{\brr}{\begin{array}}
 
\newcommand{\err}{\end{array}}
\newcommand{\bc}{\begin{center}}
\newcommand{\ec}{\end{center}}

\newcommand{\tu}{\textunderscore}

\DeclareMathAlphabet{\mathsc}{OT1}{cmr}{m}{sc}
\def\testbx{bx}%
\DeclareRobustCommand{\ion}[2]{%
\relax\ifmmode
\ifx\testbx\f@series
{\mathbf{#1\,\mathsc{#2}}}\else
{\mathrm{#1\,\mathsc{#2}}}\fi
\else\textup{#1\,{\mdseries\textsc{#2}}}%
\fi}

\title[The SFRF of $z \sim 1-4$ galaxies]{The evolution of the star formation rate function and cosmic star formation rate density of galaxies at ${\bf{z\sim1-4}}$}

 \author[A. Katsianis et al.]{A. Katsianis$^{1,3}$\thanks{E-mail:
     kata@das.uchile.cl }, E. Tescari$^{2,3}$, G. Blanc$^1$ and M. Sargent$^4$\\  \\ $^1$ Department of Astronomy, Universitad de Chile, Camino El Observatorio 1515, Las Condes, Santiago, Chile  \\ $^2$ School of Physics, The University of Melbourne, Parkville, VIC 3010, Australia \\ $^3$ ARC Centre of Excellence for All-Sky Astrophysics (CAASTRO) \\ $^4$ Astronomy Centre, Department of Physics \& Astronomy, University of Sussex, Brighton BN1 9QH, UK}

\begin{document}

\maketitle

\begin{abstract}

We investigate the evolution of the galaxy Star Formation Rate Function (SFRF) and Cosmic Star Formation Rate Density (CSFRD) of $z\sim 1-4 $ galaxies, using cosmological Smoothed Particle Hydrodynamic (SPH) simulations and a compilation of UV, IR and H$\alpha$ observations. These tracers represent different populations of galaxies with the IR light being a probe of objects with high star formation rates and dust contents, while UV and H$\alpha$ observations provide a census of low star formation galaxies where mild obscuration occurs. We compare the above SFRFs with the results of SPH simulations run with the code {\small{P-GADGET3(XXL)}}. We focus on the role of feedback from Active Galactic Nuclei (AGN) and supernovae in form of galactic winds. The AGN feedback prescription that we use decreases the simulated CSFRD at $z < 3$ but is not sufficient to reproduce the observed evolution at higher redshifts.  We explore different wind models and find that the key factor for reproducing the evolution of the observed SFRF and CSFRD at $z \sim1-4$ is the presence of a feedback prescription that is prominent at high redshifts ($z \ge 4$) and becomes less efficient with time. We show that variable galactic winds which are efficient at decreasing the SFRs of low mass objects are quite successful in reproducing the observables.

\end{abstract}


\section{Introduction}

The Star Formation Rates (SFR) of galaxies represent a fundamental constrain for galaxy formation models.  The basic idea for calculating the average SFR of an object is to estimate the number of young bright stars with a certain age. However, in most cases, especially at high redshifts, galaxies are not spatially resolved and there is only access to their integrated spectrum. Hence, to quantify the star formation rates of galaxies, we typically rely on the observed luminosities and luminosity functions \citep{Madau2014}. Some typical star formation rate indicators are the following,

\begin{itemize}

\item Ultra-Violet (UV) luminosity:  The main advantage of the UV luminosity is that it gives a direct estimate of the young stellar population since both O and B stars are brighter in the UV than at longer wavelengths. Furthermore, at high  redshifts ($z \ge 4$) only the UV-emission from galaxies is observable with the current instrumentation. The simplest method of obtaining the SFR of an object is to assume a linear scaling between the SFR and the continuum luminosity integrated over a fixed band in the blue or near-ultraviolet. The optimal wavelength range is $1500-2800$ \AA \citep{Ken98a,smit12}. Evolutionary synthesis models provide relations between the SFR per unit mass, luminosity and the integrated color of the population. The conversion between the UV luminosity and SFR \citep{Ken98a,smit12} is found from these models to be
\begin{eqnarray}
\label{eq:SFRpara3}
{\rm SFR}_{\rm UV}  \, ({\rm M}_{\rm \odot} \, {\rm yr^{-1}}) = 1.4 \times 10^{-28} \, \, {\rm L}_{\rm UV}\, ({\rm ergs} \,s^{-1} \, {\rm Hz}^{-1})\,
\end{eqnarray}
where ${\rm L}_{\rm UV}$ is the UV luminosity of galaxies. The relation is valid from 1500 \AA to 2800 \AA and assumes a \citet{salpeter55} IMF. From $z \sim0.5$ to $z \sim3$, the majority of star formation took place in obscured and dusty environments and most of the UV photons were reprocessed by dust into IR emission \citep{LeFloch,Dole2006,Ru2011}. Therefore a dust correction is required. \\

\item  H$\alpha$ luminosity and nebular emission lines: The other SFR indicators to be discussed in this paper rely on measuring light from young, massive stars that has been reprocessed by interstellar gas or dust.  O and B stars produce large amounts of UV photons that ionise the surrounding gas. Hydrogen recombination produces line emission, including the Balmer series lines like H$\alpha$ ($6562.8 $ \AA) and H$\beta$ ($4861.2$ \AA). Other nebular emission lines from other elements, like ${\rm [O \, II]}$ \citep{Kewley2004} and ${\rm [O \, III]}$ \citep{Teplitz2000,Moustakas2006} can be used to infer the number of blue massive stars ($[O \, III]$ lines are known to be very sensitive to the ionization parameter so ${\rm [O \, II]}$ is typically preferred between them as a SFR indicator). Probing the existence of massive stars using the H$\alpha$ luminosity of an object is quite common in the literature \citep{Ken83,Gallego1995,Ken98a,Pettini1998,Glazebrook1999,Moorwood2000,Hopkins2000,Sullivan2000,Tresse2002,Pg2003,Yan2003,Hanish2006,Bell07,Ly11,Sobral2013}, since H$\alpha$ photons originate from the gas ionized by the radiation of these stars. Typically, these lines trace stars with masses greater than $ \sim$ $15M_{\odot}$, with the peak contribution from stars in the range 30-40 $M_{\odot}$. According to the synthesis models of \citet{kennicutt1998}, the relation between SFR and H$\alpha$ luminosity is the following:
\begin{eqnarray}
\label{eq:SFRpara1}
{\rm SFR}_{\rm H\alpha} \, ({\rm M}_{\rm \odot} \, yr^{-1} )= 7.9 \times 10^{-42} \, \, {\rm L}_{\rm H\alpha}\, (ergs \, s^{-1})\, ,
\end{eqnarray}
where ${\rm L}_{\rm H\alpha}$ is the H$\alpha$ luminosity of the galaxies. While it is desirable to extend H$\alpha$ studies to higher redshifts, such task is typically observationally difficult because most of the H$\alpha$ luminosity is redshifted into the infrared beyond $z \sim0.4$. \\

\item The Infra-Red (IR) luminosity originating from dust continuum emission is a star formation indicator and a good test of dust physics  \citep{Hira03}. The shape of the thermal IR light depends on a lot of parameters \citep{Draine2007} like the dust opacity index, dust temperature, strength of the interstellar radiation field and Polycylic Aromatic Hydrocarbons (PAH) abundance, in the sense that UV-luminous, young stars will heat the dust to higher temperatures than older stellar populations \citep{Helou1986}. The dust (heated by UV-luminous and young stellar populations) produce an IR spectral distribution that is more luminous than the one produced by low mass stars. This is the foundation for using the IR emission ( $\sim 5-1000 \mu m$) as a probe of UV-bright stars and a SFR indicator. There are two approaches to study the SFR using IR observations. Both involve IR photometry as a tracer of IR luminosity, which in turn traces the number of UV photons from the short-lived massive stars and allows the SFR to be calculated \citep{Ken98a,Ru2011}. The first approach involves multi-band IR photometry and constrains the total IR luminosity \citep{Elbaz10,Rex10}, while the second exploits just a monochromatic IR luminosity \citep[][IR at 24 $\mu$m]{Calzetti07} that correlates strongly with SFR. The relation between the SFR and total IR luminosity from the evolutionary synthesis model of \citet{kennicutt1998} is found to be:
\begin{eqnarray}
\label{eq:SFRpara2}
{\rm SFR}_{\rm IR} \, ({\rm M}_{\rm \odot} \, yr^{-1} )= 1.72 \times 10^{-10} \, \, {\rm L}_{\rm IR} \,/{\rm L}_{\rm \odot}\,
\end{eqnarray}

\end{itemize} 

There has been a considerable effort to constrain the evolution of the cosmic star formation rate density in the last decade\citep{Madau2014}. However, \citet{Ly11} state that it is important to explore the extent to which systematics between different SFR indicators can affect its measurements. This can be done by comparing the SFRFs obtained from different indicators. In particular, it is useful to trace the star formation history with a single indicator throughout time, and then compare the overall histories from various SFR tracers. 

The evolution of the SFRF has been studied by means of hydrodynamic simulations \citep{Dave2011,TescariKaW2013} and semi-analytic modelling \citep{Fontanot2012}.  \citet{Dave2011} used a set of simulations run with an improved version of GADGET-2 to study the growth of galaxies from $z \sim0-3$. The authors investigated the effect of four different wind models and compared the simulated star formation rate functions with observations \citep{Martin05,Hayes2010,Ly11}. Their galactic wind models are responsible for the shape of the faint end slope of the SFR function at $z=0$.  However, the simulations overproduce the number of objects at all SFRs. According to the authors, this tension is due to the absence of Active Galactic Nuclei (AGN) feedback in their models.

This paper is the fourth of a series in which we present the results of the {\textsc{Angus}} ({\textit{AustraliaN {\small{GADGET-3}} early Universe Simulations}}) project and the observed SFRF of $z \sim1-4$ galaxies, that were obtained from a compilation of UV, H$\alpha$ and IR Luminosity Functions (LF). The aim of the {\textsc{Angus} project is to study the interplay between galaxies and the Intergalactic Medium from intermediate redshifts ($z \sim1$) to the epoch of reionization at $z \sim6$ and above. We use the hydrodynamic code {\small{P-GADGET3(XXL)}}, which is an improved version of {\small{GADGET-3}} \citep{Springel2005}. For the first time we combine physical processes, which have been developed and tested separately. In particular, our code includes:
\begin{itemize}
\item a sub-grid star formation model \citep{springel2003},
\item supernova energy- and momentum-driven galactic winds \citep{springel2003,PuchweinSpri12,barai13},
\item AGN feedback \citep{springeletal05nat,Fabjan,susana13},
\item self-consistent stellar evolution and chemical enrichment modeling \citep{tornatore07},
\item metal-line cooling \citep{wiersma09},
\item transition of metal free Population III to Population II star formation \citep{tornatore07b},
\item a low-viscosity SPH scheme to allow the development of turbulence within the intracluster medium \citep{dolag2005},
\item low-temperature cooling by molecules/metals \citep{maio2007},
\item thermal conduction \citep{Dolag2004},
\item passive magnetic fields based on Euler potentials \citep{DolagSta2009},
\item adaptive gravitational softening \citep{iannuzzi11}.
\end{itemize}} Simulations based on the same code have also been used to successfully explore the origin of cosmic chemical abundances \citep{MaioTe2015}. In \citet{TescariKaW2013} we constrained and compared our numerical results  with observations of the Star Formation Rate Function (SFRF) at $z \sim4-7$ \citep{smit12}. In addition, we showed that a fiducial model with strong-energy driven winds and AGN feedback which starts to be effective at high redshifts ($z \ge 4$) is needed to obtain the observed SFRF of high redshift galaxies. In this work we extend the analysis to lower redshifts ($1\le z\le 4$) using the same set of cosmological simulations. We explore various feedback prescriptions and investigate how these shape the galaxy SFRF. We do not investigate the broad possible range of simulations, but concentrate on the simulations that can describe the high-$z$ SFR function \citep{TescariKaW2013}, Galaxy Stellar Mass Function (GSMF) and  SFR$-{\rm M}_{\star}$ relations \citep{Katsianis2014,Katsianis2015}. 

This paper is organized as follows. In section \ref{SFRindi} we present the compilation of the observed luminosity functions and dust correction laws used for this work. In Section \ref{Obsout6} we present the observed SFRF of galaxies at $z \sim 1-4$. In Section \ref{thecode} we present a brief description of our simulations along with the different feedback models used. In section \ref{SFRFbhydro6} we compare the simulated SFRFs with the constrains from the observations. In section \ref{CSFRDsim} we present the evolution of the cosmic star formation rate density of the Universe in observations and simulations. Finally, in Section \ref{concl6} we summarise our main results and conclusions.

\section{The observed star formation rates from galaxy Luminosities}
\label{SFRindi}

\subsection{Dust attenuation effects and dust correction prescriptions}
\label{dustcorrectionlaws}

\begin{table}
\centering
\begin{tabular}{llccc}
  \\ \hline & SFR Indicator & Dust corrections \\ \hline
  & \textit{UV}  & \citet{meurer1999} and \citet{HaoKen}  \\
  & \textit{IR}  & No dust corrections needed \\
  & \textit{H$\alpha$} &  1 mag, \citet{Hopkins01}  \\
  \hline \\
\end{tabular}
\caption{The dust correction formulas used for the UV, IR and H$\alpha$ luminosities in this work.}
\label{CorrectLaws}
\end{table}

We correct the UV LFs for the effects of dust attenuation  using the correlation of extinction with the UV-continuum slope $\beta$ following \citet{HaoKen} and \citet{smit12}. Like \citet{smit12} we assume the infrared excess (IRX)-$\beta$ relation of \citet{meurer1999}:
\begin{eqnarray}
  A_{\rm 1600} = 4.43 + 1.99\,\beta,
  \label{eq_A16}
\end{eqnarray}
where $A_{\rm 1600}$ is the dust absorption at 1600 \AA. We
assume as well the linear relation between the UV-continuum slope $\beta$ and
luminosity of \citet{bouwens2012}:
\begin{eqnarray}
  \langle\beta\rangle=\frac{{\rm d}\beta}{{\rm d}M_{\rm UV}}\left(M_{\rm
      UV,AB}+19.5\right)+\beta_{M_{\rm UV}=-19.5},
  \label{eq_beta}
\end{eqnarray}
Then following \citet{HaoKen} we assume
\begin{eqnarray}
  {\rm L_{\rm UV_{OBS}}} = {\rm L_{\rm UV_{corr}}e^{-\tau_{UV}}},
  \label{eq_A17}
\end{eqnarray}
where ${\rm \tau_{UV}}$ is the effective optical depth (${\rm \tau_{UV}}={\rm A_{\rm 1600}/1.086}$). We calculate $A_{\rm 1600}$ and ${\rm \tau_{UV}}$ adopting the parameters for $\frac{{\rm d}\beta}{{\rm d}M_{\rm UV}}$ from \citet{Reddy2009}, \citet{bouwens09,bouwens2012} and \citet{Tacchella2013}. 

For the case of H$\alpha$ emission, \cite{Sobral2013} used the 1 mag correction which is a simplification that normally is acceptable for low redshifts ($0.0<z<0.3$). \cite{Ly11} use the SFR dependent dust correction from \cite{Hopkins01}. As mentioned above, no dust corrections are required for IR luminosity functions.

\subsection{The observed UV, IR and H$\alpha$ luminosity functions from $\boldsymbol{z \sim3.8}$ to $\boldsymbol{z \sim0.8}$}
\label{intr61}

To retrieve the SFRF for redshift $z \sim 3.8$ to $z \sim 0.8$ we use the luminosity functions from \citet[bolometric-UV+IR,][]{Reddy2008}, \citet[Lyman-break selected,][]{VAN2010}, \citet[Lyman-break selected,][]{Oesch2010}, \citet[H$\alpha$-selected,][]{Ly11},  \citet[I band selected-flux limitted,][]{cucciati12}, \citet[IR-selected,][]{Gruppionis13}, \citet[IR-selected,][]{Magenlli11,Magnelli2013}, \citet[H$\alpha$-selected,][]{Sobral2013}, \citet[Lyman-break selected,][]{Alavi13} and \citet[Lyman-break selected][]{Parsa2015}. In addition, we compare our results with the work of \citet{smit12} for Lyman-break selected galaxies at redshift $z \sim 3.8$. We choose the above surveys since all of them combined are ideal to study the SFRF in a large range of SFRs and redshifts. The authors have publicly available the LFs of their samples which are summarized below.  \\

\noindent
\citet{Reddy2008} used a sample of Lyman-break selected galaxies at redshifts $z \sim 2.3$ and $z \sim 3.1$, combined with ground-based spectroscopic H$\alpha$ and Spitzer MIPS 24 $\mu$m data, and obtained robust measurements of the rest-frame UV, H$\alpha$, and IR luminosity functions. These luminosity functions were corrected for incompleteness and dust attenuation effects. The stepwise bolometric luminosity function of \citet{Reddy2008} is in table 9 of their work.\\

\noindent
\citet{VAN2010} studied $\sim$ 100000 Lyman-break galaxies from the CFHT Legacy Survey at $z$ $\sim$ $3.1$, $3.8$, $4.8$ and estimated their rest-frame 1600 \AA \, luminosity function. Due to the large survey volume, the authors state that cosmic variance had a negligible impact on their determination of the UV luminosity function, allowing them to study the bright end with great statistical accuracy. They obtained the rest-frame UV luminosity function in absolute magnitudes at  1600 \AA \, for redshifts $z$ $\sim$ $3.1$, $3.8$, $4.8$ and their results are in table 1 of their work.\\

\noindent
\citet{Oesch2010} investigated the evolution of the UV LF at $z \sim0.75-2.5$. The authors suggested that UV-color and photometric selection have similar results for the luminosity function in this redshift interval and claim that UV-dropout samples are well defined and reasonably complete. The authors note that the characteristic luminosity decreased by a factor of $ \sim 16$ from $z \sim3$ to $z \sim0$ while the faint-end slope $\alpha$ increased from $\alpha \sim-1.5$ to $\alpha \sim-1.2$. The parameters of the analytic expressions of the above UV luminosity functions are provided in table 1 of \citet{Oesch2010}. \\

\noindent
\citet{Ly11} obtained measurements of the H$\alpha$ luminosity function for galaxies at $z \sim0.8$, based on 1.18$\mu$m narrowband imaging from the NewH$\alpha$ Survey. The authors applied corrections for dust attenuation effects and incompleteness. To correct for dust attenuation, they adopted a luminosity-dependent extinction relation following \citet{Hopkins01}. The authors applied corrections for [N II] flux contamination and the volume, as a function of line flux. The H$\alpha$ luminosity function from \citet{Ly11} is presented in table 3 of their work.\\

\noindent
\citet{cucciati12} investigated the evolution of the far ultraviolet (FUV) and near ultraviolet (NUV) luminosity functions from $z \sim 0.05$ to $z \sim 4.5$. Using these data, they derived the CSFRD history and suggested that it peaks at $z \sim2$ as it increases by a factor of $\sim 6$ from $z \sim4.5$. We use the FUV luminosity functions to obtain an estimate of the SFRs of small galaxies. The analytic expressions of the above FUV luminosity functions are provided in table 1 of \citet{cucciati12}.\\

\noindent
\citet{Gruppionis13} used the 70-, 100-, 160-, 250-, 350- and 500-$\mu$m data from the  Herschel surveys, PEP and HerMES, in the GOODS-S and -N, ECDFS and COSMOS fields, to characterise the evolution of the IR luminosity function for redshifts $z \sim4$ to $z \sim0$. The highest redshift results of \citet{Gruppionis13} provide some information on the bright end of the luminosity function for high redshifts. The authors provide more useful constrains at lower redshift where IR luminosity functions can probe low star forming objects. The total IR luminosity function of \citet{Gruppionis13} is in table 6 of their work. \\ 

\noindent
\citet{Magenlli11,Magnelli2013} combined observations of the GOODS fields from the PEP and GOODS-Herschel programmes. From the catalogues of these fields they derived number counts and obtained the IR luminosity functions down to $L_{IR}=10^{11} \, L_{\odot}$ at $z \sim1$ and  $L_{IR}=10^{12} \, L_{\odot}$ at $z \sim2$. The authors state that their far-infrared observations provide a  more accurate infrared luminosity estimation than the mid-infrared observations from Spitzer. The results of \citet{Magnelli2013} are presented in the appendix of their work. \\

\noindent
\citet{Sobral2013} presented the combination of  wide and deep narrow-band H$\alpha$ surveys using UKIRT, Subaru and the VLT. The authors robustly selected a total of 1742, 637, 515 and 807 H$\alpha$ emitters across the COSMOS and the UDS fields at z = 0.40, 0.84, 1.47 and 2.23, respectively. These H$\alpha$ LFs have then been corrected for incompleteness, [NII] contamination and for dust extinction using $A_{H\alpha}$ = 1 mag correction\footnote{ \citet{Hopkins01} note that a SFR-dependent dust attenuation law produces similar corrections with the 1 magnitude simplification often assumed for local populations. At higher redshifts ($z > 0.3$) though, larger corrections are required.}. The stepwise determination of the H$\alpha$ luminosity function from \citet{Sobral2013} can be found in table 4 of their work.\\

\noindent
\citet{Alavi13} targeted the cluster Abell 1689, behind which they searched for faint star-forming galaxies at redshift $z \sim2$. Their data are corrected for  incompleteness and dust attenuation effects. They extended the UV luminosity function at $z \sim2$ to very faint magnitude limits and this allowed them to constrain the $\alpha$ parameter of the Schechter function fit, finding that $\alpha=-1.74 \, \pm 0.08$. The parameters of the UV luminosity function found with this method are in table 3 of \citet{Alavi13}. \\

\noindent
\citet{Parsa2015} present measurements of the evolving rest-frame UV (1500 \AA) galaxy luminosity  function over the redshift range $z \sim2-4$. The results are provided by combining the HUDF, CANDELS/GOODS-South, and UltraVISTA/COSMOS surveys and are able to succesfully probe the faint end of the UV luminosity function. An interesting result of the analysis is that the LF appears to be significantly shallower ($ \alpha = -1.32$) than previous measurements \citep[e.g.][]{Alavi13}. The stepwise determination of the LF found by this method is in table 1 of \citet{Parsa2015}. 

\subsection{The luminosity-SFR conversion}
\label{LSFR62}

To obtain the SFRF of galaxies we start from the observed luminosity function. We convert the luminosities to SFRs at each bin of the LF following a method similar to the one adopted by \citet{smit12}. UV selected samples provide information about the SFRF of galaxies at redshifts $z > 2$. For lower redshifts they provide key constrains for low star forming objects (faint-end of the distribution), but are unable to probe dusty high star forming systems, and thus are uncertain at the bright-end of the distribution. This is due to the fact that dust corrections are insufficient or that UV selected samples do not include a significant number of massive, dusty objects. Using IR selected samples we obtain SFRFs that are not affected by dust attenuation effects. However, small faint galaxies do not have enough dust to reprocess the UV light to IR, so IR luminosities do not probe the faint end of the SFRF. The observed H$\alpha$ data used for this work provide us with information about the SFRF of intermediate $ 1.0 \leq \log(SFR/ (\, M_{\odot} \, yr^{-1}))  \leq  2.0 $ galaxies from $z \sim0.8$ to $z \sim2.3$. Dust corrections are required to obtain the intrinsic SFRs from H$\alpha$ luminosities. We note that various authors have often published their data in the form of UV, IR or H$\alpha$ luminosity functions. However, the authors used their results to directly obtain the cosmic star formation rate density instead of star formation rate functions since the first is known to be dominated mostly by galaxies around the characteristic luminosity. The different groups are aware of the potential problems (e.g. uncertainty in treatment of dust) that may occur when they calculate SFRs from a range of individual galaxy luminosities or bins of LFs.

\section{The observed star formation rate functions from $z \sim3.8$ to $z \sim0.8$}
\label{Obsout6}

\begin{table}
  \centering
  \begin{tabular}{cccc}
    \hline \\
    & {\large $\log\,\frac{{\rm SFR}}{{\rm M}_{\odot}\ {\rm yr}^{-1}}$}  &
    {\large $\log \, \phi_{\rm SFR}\ \left({\rm Mpc}^{-3}\ {\rm
          dex}^{-1}\right)$} \\ \\
    \hline \hline
    & $z\sim3.8$, \citep[][ UV]{Parsa2015} & dust corrected \\ 
    \hline 
    &-0.86  & -1.57$\pm$0.03  \\  
    &-0.66  & -1.68$\pm$0.04  \\ 
    &-0.47  & -1.49$\pm$0.03  \\ 
    &-0.26  & -1.83$\pm$0.05  \\ 
    &-0.02  & -1.96$\pm$0.06  \\ 
    & 0.23  & -2.10$\pm$0.07  \\ 
    & 0.47  & -2.37$\pm$0.10  \\ 
    & 0.72  & -2.42$\pm$0.28  \\ 
    & 0.96  & -2.47$\pm$0.03  \\
    & 1.20  & -2.66$\pm$0.04  \\
    & 1.45  & -3.02$\pm$0.01  \\ 
    & 1.69  & -3.51$\pm$0.03  \\
    & 1.93  & -4.07$\pm$0.05  \\
    & 2.18  & -4.70$\pm$0.09  \\
    & 2.42  & -5.70$\pm$0.43  \\  
    \hline
    & $<z> \sim3.6$, \citep[][ IR]{Gruppionis13} & No correction \\ 
    \hline 
    & 2.49  & -4.65$\pm$0.14  \\
    & 2.99  & -5.75$\pm$0.13 \\
    & 3.48  & -7.18$\pm$0.43  \\
    \hline 
    & $z\sim3.8$, \citep[][ UV]{VAN2010} & dust corrected \\ 
    \hline 
    & 0.33  & -1.99$\pm$0.22  \\
    & 0.47  & -2.08$\pm$0.08 \\
    & 0.62  & -2.13$\pm$0.06  \\
    & 0.76  & -2.23$\pm$0.05  \\
    & 0.91  & -2.31$\pm$0.07  \\
    & 1.06  & -2.47$\pm$0.09 \\
    & 1.20  & -2.62$\pm$0.05  \\
    & 1.35  & -2.77$\pm$0.04  \\
    & 1.50  & -3.02$\pm$0.02  \\
    & 1.64  & -3.27$\pm$0.04 \\
    & 1.79  & -3.54$\pm$0.02  \\
    & 1.93  & -3.93$\pm$0.06  \\
    & 2.08  & -4.27$\pm$0.10  \\
    & 2.23  & -4.88$\pm$0.05 \\
    \hline \\
  \end{tabular}
  \caption{Stepwise SFR functions at $z\sim3.8$ using the data from \citet[][ blue triangles in the top left panel of Fig. \ref{fig:38SFRF}]{VAN2010}, \citet[][ red diamonds in the top left panel of Fig. \ref{fig:38SFRF}]{Gruppionis13} and \citet[][ grey squares in the top left panel of Fig. \ref{fig:38SFRF}]{Parsa2015}.   }
  \label{tab_stepsfrf4}
\end{table}

\subsection{The Star formation rate function at $\boldsymbol{z \sim3.8}$}
\label{SFRF380}

To obtain the SFRF at $z \sim4$ we start with the UV luminosity function from \citet[][ $z \sim3.8$]{VAN2010}. We use the dust corrections laws of \citet{meurer1999} and \citet{HaoKen} and obtain the dust corrected UV luminosity functions. We assume the same  $< \beta >$ as \citet{smit12}  that can be found in \citet[][ $< \beta > = -0.11(M_{UV, AB}+19.5)-2.00$ at $z \sim3.8$]{bouwens2012}. We apply equation \ref{eq:SFRpara3} \citep{kennicutt1998,madau1998}, which assumes a \citet{salpeter55} IMF and convert the UV luminosities to SFRs. The blue triangles in the top left panel of Fig. \ref{fig:38SFRF}  are the stepwise determinations of the SFRF for redshift $z \sim 3.8$ using the above method. The red triangles are the stepwise determination  of the SFRF for $z \sim3.8$ from \citet{smit12}.  The results from our analysis and those of \citet{smit12} imply that the UV luminosity functions from \citet{bouwens2007} and  \citet{VAN2010} are consistent with each other for $z \sim4$. The error bars of the SFRF that relies on the results of \citet{VAN2010} are smaller, due to less  cosmic variance within the larger area covered by these observations. We use the same method to convert the UV luminosity function of \citet[][ $z \sim 3.8$]{Parsa2015} to a stepwise SFRF. However, the SFRF that was obtained from the UV LF of \citet{Parsa2015} has a shallower faint end. The results of \citet{Parsa2015} include fainter sources and thus provide further information for low star forming objects. The SFRFs from our analysis for $z \sim3.8$ are in the top left panel of Fig. \ref{fig:38SFRF} and Table \ref{tab_stepsfrf4}. In addition, we use the IR luminosity function from \citet{Gruppionis13} for redshifts $3.0 < z <4.2$. We convert the IR luminosities to SFRs using equation \ref{eq:SFRpara2}. We see that the IR luminosity is unable to provide information for a broad population of $z \sim 4$ galaxies and is only a census of objects with very high SFRs. Thus, unlike UV LFs, rest frame IR studies are not suitable for constraining the star formation rate function and CSFRD at high redshifts.

\begin{table}
  \centering
\resizebox{0.50\textwidth}{!}{%
  \begin{tabular}{cccc}
    \hline \\
    & {\large $\log\,\frac{{\rm SFR}}{{\rm M}_{\odot}\ {\rm yr}^{-1}}$}  &
    {\large $\log \, \phi_{\rm SFR}\ \left({\rm Mpc}^{-3}\ {\rm
          dex}^{-1}\right)$} \\ \\
    \hline \hline
    & $z\sim3.0$, \citet[][ UV]{Parsa2015} & dust corrected \\ 
    \hline
    &-1.06  & -1.54$\pm$0.03  \\
    &-0.86  & -1.51$\pm$0.03  \\
    &-0.66  & -1.44$\pm$0.03  \\
    &-0.42  & -1.61$\pm$0.04  \\
    &-0.17  & -1.60$\pm$0.04  \\ 
    & 0.08  & -1.77$\pm$0.05  \\
    & 0.33  & -1.83$\pm$0.05  \\
    & 0.58  & -2.06$\pm$0.02  \\
    & 0.84  & -2.12$\pm$0.02  \\
    & 1.09  & -2.30$\pm$0.02  \\
    & 1.39  & -2.53$\pm$0.03  \\
    & 1.60  & -3.03$\pm$0.01  \\
    & 1.84  & -3.63$\pm$0.03  \\
    & 2.09  & -4.34$\pm$0.08  \\
    & 2.35  & -5.22$\pm$0.14  \\
    \hline
    & $z\sim3.1$, \citet[][ UV]{VAN2010} & dust corrected \\ 
    \hline 
    & 0.43  & -1.97$\pm$0.20  \\
    & 0.58  & -1.80$\pm$0.08 \\
    & 0.73  & -1.91$\pm$0.10  \\
    & 0.89  & -2.03$\pm$0.08  \\
    & 1.04  & -2.13$\pm$0.07  \\
    & 1.19  & -2.24$\pm$0.09 \\
    & 1.34  & -2.40$\pm$0.10  \\
    & 1.49  & -2.58$\pm$0.06  \\
    & 1.64  & -2.78$\pm$0.05  \\
    & 1.79  & -3.07$\pm$0.05 \\
    & 1.94  & -3.42$\pm$0.11  \\
    & 2.09  & -3.72$\pm$0.15  \\
    & 2.25  & -4.13$\pm$0.14  \\
    & 2.40  & -4.63$\pm$0.12 \\
    & 2.55  & -5.48$\pm$0.43  \\
    & 2.70  & -5.48$\pm$0.43 \\
    \hline
     & $z\sim3.1$, \citet[][ Bolometric]{Reddy2008}  & dust corrected \\ 
    \hline 
    &-0.14  & -1.59$\pm$0.27  \\
    & 0.09  & -1.69$\pm$0.24 \\
    & 0.36  & -1.83$\pm$0.24  \\
    & 0.61  & -2.02$\pm$0.27  \\
    & 0.86  & -2.22$\pm$0.26  \\
    & 1.11  & -2.41$\pm$0.21 \\
    & 1.36  & -2.60$\pm$0.18  \\
    & 1.61  & -2.83$\pm$0.13  \\
    & 1.86  & -3.11$\pm$0.11  \\
    & 2.11  & -3.40$\pm$0.08 \\
    & 2.36  & -3.73$\pm$0.12  \\
    & 2.61  & -4.12$\pm$0.12  \\
   \hline
  \end{tabular}%
}
  \caption{Stepwise SFR functions at $z\sim3.1$ using the data from  \citet{Reddy2008},  orange circles in the top right panel of Fig. \ref{fig:38SFRF}, \citet{VAN2010}, blue triangles in the top right panel of Fig. \ref{fig:38SFRF} and \citet{Parsa2015}, grey squares in the top right panel of Fig. \ref{fig:38SFRF}.   }
  \label{tab_stepsfrf3}
\end{table}

\subsection{The Star formation rate function at $\boldsymbol{z \sim3.1}$}
\label{SFRF310}

To obtain the SFRF at redshift $z \sim3.1$ we start with the UV luminosity function from \citet[][ $z \sim3.1$]{VAN2010} and \citet[][ $z \sim3.0$]{Parsa2015}. We follow the same procedure described above. We assume the $< \beta >$ relations from \citet{Reddy2009} that can be found in \citet{Tacchella2013} who show $< \beta > = -0.13(M_{UV, AB}+19.5)-1.85$ at $z \sim3.0$. The blue triangles in the top right panel of Fig. \ref{fig:38SFRF} are the stepwise determination of the SFRF for redshift $z \sim 3.1$ using the UV luminosity function of  \citet{VAN2010}, while the grey squares are the SFRF from \citet{Parsa2015}. In addition, we use the bolometric luminosity function from \citet[][ $z \sim 2.7-3.4$]{Reddy2008}. This luminosity function is already corrected for incompleteness and dust attenuation effects. To obtain the SFRF from the bolometric luminosity function we use equation \ref{eq:SFRpara2}. The orange filled circles in the top right panel of Fig. \ref{fig:38SFRF} are the stepwise determination of the SFRF for $z \sim3.1$ using the above procedure.

We see that the SFRFs we obtain using the data of \citet{Reddy2008}, \citet{VAN2010} and \citet{Parsa2015} are in good agreement. The SFRF we retrieve employing the LF of \citet{Parsa2015} can probe objects with low star formation rates and thus provide better constraints for the small objects in our simulations. The red diamonds are obtained using the IR data of \citet{Gruppionis13} for $z \sim2.5-3.0$ and will be discussed in detail in the next Section. The results for $z \sim3.1$ are sumarized in Table \ref{tab_stepsfrf3} and the top right panel of Fig. \ref{fig:38SFRF}.

\begin{table}
  \centering
\resizebox{0.50\textwidth}{!}{%
  \begin{tabular}{cccc}
    \hline \\
    & {\large $\log\,\frac{{\rm SFR}}{{\rm M}_{\odot}\ {\rm yr}^{-1}}$}  &
    {\large $\log \, \phi_{\rm SFR}\ \left({\rm Mpc}^{-3}\ {\rm
          dex}^{-1}\right)$} \\ \\
    \hline \hline
     & $z\sim2.5-3.0$, \citet[][ IR]{Gruppionis13}  & No correction \\ 
    \hline 
    & 2.49  & -3.75$\pm$0.21  \\
    & 2.99  & -4.15$\pm$0.11 \\
    & 3.49  & -5.11$\pm$0.07  \\
    \hline \\
  \end{tabular}%
}
  \caption{Stepwise determinations of the SFR function at $z\sim2.6$  (red diamonds Fig. \ref{fig:38SFRF}) using the IR luminosity function of \citet{Gruppionis13} for $2.5 < z < 3.0$. The parameters of the analytic expression (dark green dotted line in Fig. \ref{fig:38SFRF}) we obtain from dust correcting the \citet{schechter1976} fit given by \citet{Oesch2010} are: $\phi^{\star} = 0.0026$ ${\rm Mpc}^{-3}$, $SFR^{\star}=24.56$ $M_{\odot}/yr $ and $\alpha = -1.48$. }
  \label{tab_stepsfrf26}
\end{table}

\subsection{The Star formation rate function at $\boldsymbol{z \sim2.6}$}
\label{SFRF260}

\begin{figure*}
\centering
\includegraphics[scale=0.83]{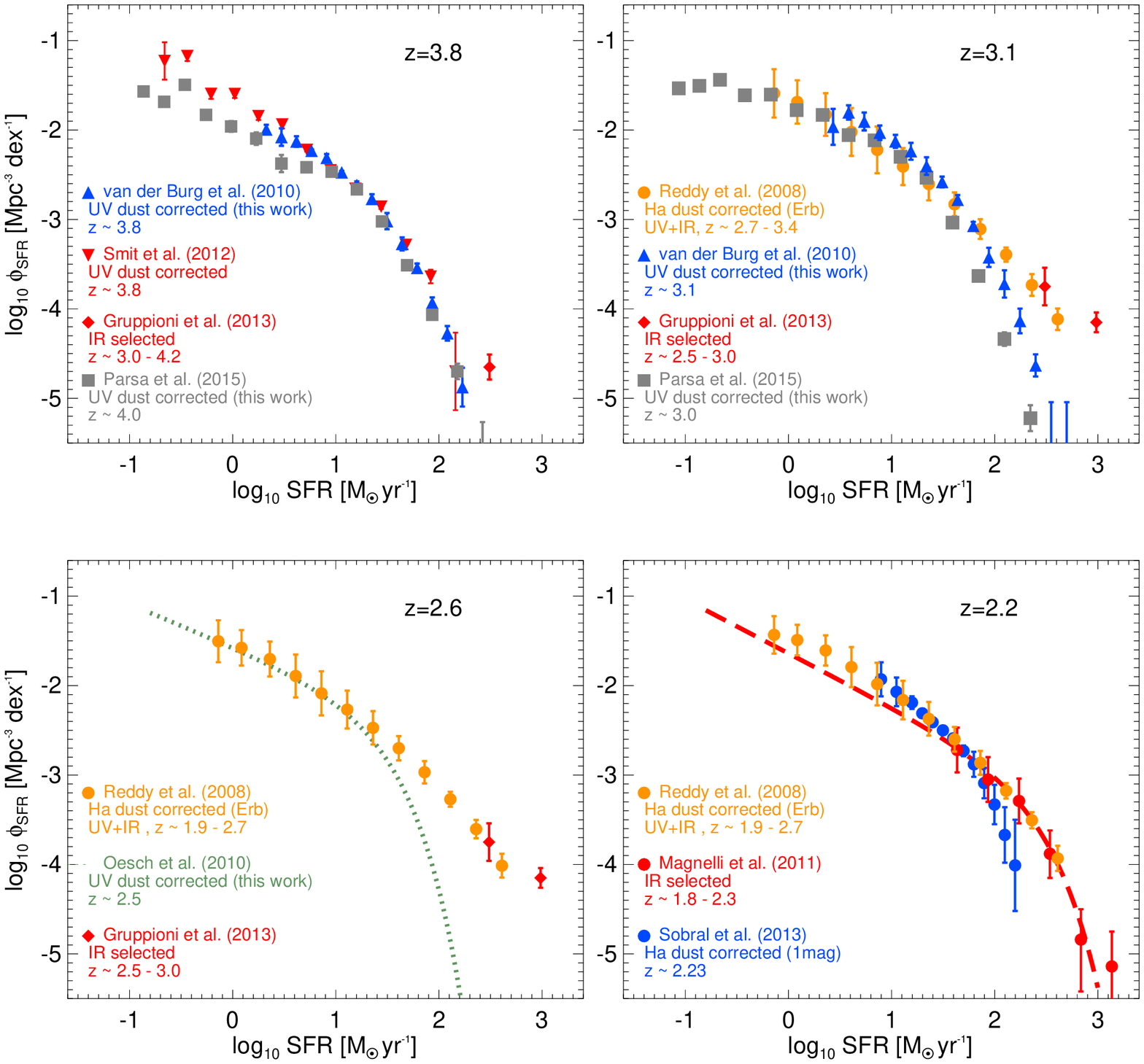}
\vspace{0.05cm}
\caption{The stepwise and analytical determinations of the observed SFRF for redshifts $z \sim 3.8$ (top left panel), $z \sim 3.1$ (top right panel), $z \sim 2.6$ (bottom left panel) and $z \sim 2.2$ (bottom right panel). The blue filled circles are the SFRFs for $z \sim 2.2$ using the H$\alpha$ luminosity function from \citet{Sobral2013}. The red filled circles and red dashed-line are the SFRFs for redshift $z \sim 2.2$ using the IR luminosity function from \citet{Magenlli11}.  The red filled diamonds are the SFRF for redshifts $z \sim 2.6$, $z \sim3.1$ and $z \sim3.8$ using the IR luminosity function from \citet{Gruppionis13}. The orange filled circles are the SFRF for redshifts $z \sim 2.2$, $z \sim 2.6$ and $z \sim 3.1$ using the bolometric luminosity function from \citet{Reddy2008}. The green dotted line is the analytic SFRF obtained using the results of \citet{Oesch2010}. The blue triangles at  $z \sim 3.1$ and $z \sim 3.8$ were retrieved from dust correcting the UV luminosity functions of  \citet{VAN2010}. For the grey squares we used the UV luminosity function from  \citet{Parsa2015}. The red inverted triangles are the stepwise determination  of the SFRF at $z \sim3.8$ from \citet{smit12}.}
\label{fig:38SFRF}
\end{figure*}

To obtain the SFRF for $z \sim2.6$ galaxies we use the bolometric data of \citet{Reddy2008} at $z \sim1.9-3.4$, the Schechter fit of the UV luminosity function of \citet[][ $z\sim 2.5$]{Oesch2010} and the IR LF of \citet{Gruppionis13} at $z\sim 2.5-3.0$. To convert the LF of \citet{Oesch2010} we assume the $< \beta >$ relations from \citet{bouwens09,bouwens2012} who find $< \beta > = -0.20(M_{UV, AB}+19.5)-1.70$ at $z \sim2.5$. The results of this analysis are shown in the bottom left panel of Fig. \ref{fig:38SFRF} (green dotted line). Moreover, we use equation \ref{eq:SFRpara2} to convert the IR luminosities  of \citet{Gruppionis13} to SFRs. The SFRF from the IR luminosity function is only able to constrain the SFRs of luminous star forming systems at this high redshift. However, they indicate that the UV SFRs that are obtained from IRX-$\beta$ relation are underestimated or the Lyman-break selected sample of \citet{Oesch2010}  misses a significant number of high star forming systems. The results for $z \sim2.6$ are summarized in Table \ref{tab_stepsfrf26} and the bottom left panel of Fig. \ref{fig:38SFRF}.

\begin{table}
  \centering
\resizebox{0.50\textwidth}{!}{%
  \begin{tabular}{cccc}
    \hline \\
    & {\large $\log\,\frac{{\rm SFR}}{{\rm M}_{\odot}\ {\rm yr}^{-1}}$}  &
    {\large $\log \, \phi_{\rm SFR}\ \left({\rm Mpc}^{-3}\ {\rm
          dex}^{-1}\right)$} \\ \\
    \hline \hline
    & $z\sim2.23$, \citet[][ H$\alpha$]{Sobral2013}  & dust corrected \\ 
    \hline 
    & 0.90  & -1.93$\pm$0.19  \\
    & 1.05  & -2.07$\pm$0.16 \\
    & 1.20  & -2.19$\pm$0.07  \\
    & 1.30  & -2.31$\pm$0.05  \\
    & 1.40  & -2.41$\pm$0.05  \\
    & 1.50  & -2.50$\pm$0.04 \\
    & 1.60  & -2.59$\pm$0.05  \\
    & 1.70  & -2.73$\pm$0.06  \\
    & 1.80  & -2.88$\pm$0.14  \\
    & 1.90  & -3.09$\pm$0.17 \\
    & 2.00  & -3.33$\pm$0.22  \\
    & 2.10  & -3.67$\pm$0.31  \\
    & 2.20  & -4.01$\pm$0.51  \\
    \hline
    & $z \sim2.3$, \citet[][ IR]{Magenlli11}  & dust corrected \\
    \hline 
    & 1.64  & -2.72$\pm$0.25  \\
    & 1.94  & -3.05$\pm$0.25 \\
    & 2.24  & -3.29$\pm$0.25  \\
    & 2.54  & -3.88$\pm^{0.26}_{0.27}$  \\
    & 2.84  & -4.84$\pm^{0.34}_{0.58}$ \\
    & 3.14  & -5.14$\pm^{0.39}_{3.00}$  \\
    \hline
    & $z \sim 1.8-2.3$, \citet[][ Bolometric]{Reddy2008}  & dust corrected \\
    \hline
    &-0.14  & -1.43$\pm$0.21  \\
    & 0.09  & -1.49$\pm$0.17 \\
    & 0.36  & -1.61$\pm$0.17  \\
    & 0.61  & -1.79$\pm$0.22  \\
    & 0.86  & -1.98$\pm$0.23  \\
    & 1.11  & -2.16$\pm$0.21 \\
    & 1.36  & -2.37$\pm$0.19  \\
    & 1.61  & -2.60$\pm$0.14  \\
    & 1.86  & -2.86$\pm$0.13  \\
    & 2.11  & -3.18$\pm$0.09 \\
    & 2.36  & -3.51$\pm$0.09  \\
    & 2.61  & -3.93$\pm$0.14  \\
    \hline \\
  \end{tabular}%
}  
  \caption{Stepwise SFR functions at $z\sim2.2$ using the data from \citet{Reddy2008}, orange circles in the bottom right panel of Fig. \ref{fig:38SFRF}, \citet{Magenlli11}, red circles in the top bottom right of Fig. \ref{fig:38SFRF} and \citet{Sobral2013}, blue circles in the bottom right panel of Fig. \ref{fig:38SFRF}.}
  \label{tab_stepsfrf22}
\end{table}

\subsection{The Star formation rate function at $\boldsymbol{z \sim2.2}$}
\label{SFRF220}

To obtain the SFRF at redshift $z \sim2.2$ we use the luminosity functions of \citet[][ $z\sim 1.9-2.7$]{Reddy2008}, \citet[][ $z \sim1.8-2.3$]{Magenlli11} and \citet[][ $z\sim2.23$]{Sobral2013}.  The orange filled circles in the bottom right panel of Fig. \ref{fig:38SFRF}  represent the stepwise determination of the SFRF for redshift $z \sim 2.2$ using the bolometric luminosity function at $z \sim2.3$ from \citet{Reddy2008} and equation \ref{eq:SFRpara2}. In addition, we use the H$\alpha$ luminosity function from \citet{Sobral2013}. As discussed in Subsection \ref{intr61}, this luminosity function is corrected for incompleteness and dust attenuation effects (1 mag simplification). To obtain the SFRF from the H$\alpha$ luminosity function we use equation \ref{eq:SFRpara1}. The blue filled circles of Fig. \ref{fig:38SFRF} are the stepwise determination of the SFRF for $z \sim2.2$ using the above analysis. Finally, we use the IR luminosity function of \citet{Magenlli11} for redshifts $1.8 < z <2.3$.  We convert the IR luminosities to SFRs using eq. \ref{eq:SFRpara2}. We obtain the analytic form (red dashed line) converting the analytic luminosity functions to SFRFs following \citet{smit12}.  The results of \citet{Magenlli11} span the range $\sim 40$ $M_{\odot}/yr $ to $1380$ $M_{\odot}/yr $ but we extend the \citet{schechter1976} form to lower SFRs, so we can have a comparison with the other data-sets present in this work. We note that \citet{Magenlli11} obtained the CSFRD by integrating the extended \citet{schechter1976} form of their LF (without limits) so they can compare their results with other authors. This methodology was employed by most authors, thus we extend the analytic expressions of the SFRFs to lower SFRs at all panels. The SFRFs for $z \sim2.2$ galaxies from our analysis are presented in the bottom right panel of Fig. \ref{fig:38SFRF} and Table \ref{tab_stepsfrf22}. We see an excellent agreement between the different SFRFs derived from various SFR tracers. As we have seen in the previous subsections the results from the IR luminosity function \citep{Magenlli11} provide information for high star forming systems while the bolometric luminosity of \citet{Reddy2008} spans a wider range of SFRs due to their UV data. The SFRF we obtain from the H$\alpha$ results of \citet{Sobral2013} is in good agreement with the other two. 

\begin{table}
  \centering
\resizebox{0.50\textwidth}{!}{%
  \begin{tabular}{cccc}
    \hline \\
    & {\large $\log\,\frac{{\rm SFR}}{{\rm M}_{\odot}\ {\rm yr}^{-1}}$}  &
    {\large $\log \, \phi_{\rm SFR}\ \left({\rm Mpc}^{-3}\ {\rm
          dex}^{-1}\right)$} \\ \\
    \hline \hline
    & $z\sim2.0$, \citet[][ UV]{Parsa2015} & dust corrected \\ 
    \hline
    &-1.56  & -1.17$\pm$0.02 \\ 
    &-1.26  & -1.22$\pm$0.02  \\
    &-1.06  & -1.44$\pm$0.03  \\
    &-0.81  & -1.45$\pm$0.03  \\
    &-0.56  & -1.54$\pm$0.04  \\
    &-0.30  & -1.60$\pm$0.04  \\
    &-0.05  & -1.70$\pm$0.04 \\
    & 0.20  & -1.80$\pm$0.05  \\
    & 0.45  & -1.89$\pm$0.02  \\
    & 0.70  & -2.00$\pm$0.02  \\
    & 0.95  & -2.16$\pm$0.02 \\ 
    & 1.21  & -2.48$\pm$0.03  \\
    & 1.46  & -2.95$\pm$0.14  \\
    & 1.71  & -3.52$\pm$0.27  \\
    & 1.96  & -4.17$\pm$0.57 \\
    & 2.21  & -4.62$\pm$0.98  \\ 
    \hline \\
  \end{tabular}%
}
  \caption{Stepwise SFR functions at $z\sim2.0$ using the data from \citet[][ grey squares of Fig. \ref{fig:21SFRF}]{Parsa2015}.  The parameters of the analytic expression (black dotted line) we obtain from dust correcting the \citet{schechter1976} fit given by \citet{Alavi13} are $\phi^{\star} = 0.0023$ ${\rm Mpc}^{-3}$, $SFR^{\star}=23.6$ $M_{\odot}/yr $ and $\alpha = -1.58$.  }
  \label{tab_stepsfrf20}
\end{table}

\subsection{The Star formation rate function at $\boldsymbol{z \sim2.0}$}
\label{SFRF200}

\begin{figure*}
\centering
\includegraphics[scale=0.83]{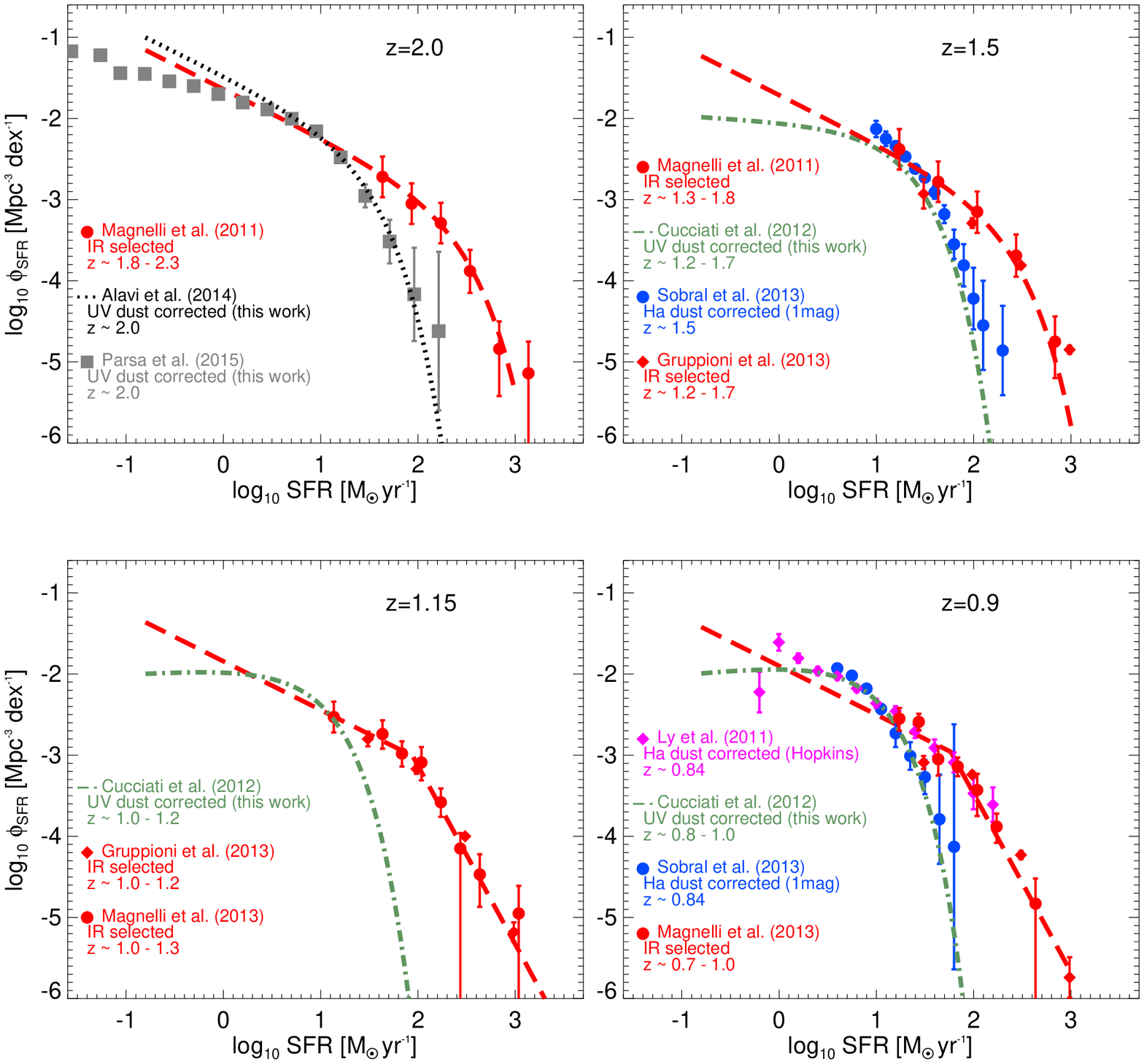}
\vspace{0.10cm}
\caption{The stepwise and analytical determinations of the observed SFRF for redshifts $z \sim 2.0$ (top left panel), $z \sim 1.5$ (top right panel), $z \sim 1.15$ (bottom left panel) and $z \sim 0.9$ (bottom right panel). The blue filled circles are the SFRFs for $z \sim 0.9$ and $z \sim1.5$ estimated using the H$\alpha$ luminosity function from \citet{Sobral2013}. The red filled circles and red dashed-lines are obtained using the IR luminosity function from \citet{Magenlli11} and \citet{Magnelli2013}. The dark green dashed-dotted lines are the analytic \citet{schechter1976} fits that we obtain after dust correcting the UV luminosity function of \citet{cucciati12}. The red filled diamonds are the SFRFs for redshifts $z \sim 0.9$, $z \sim 1.15$ and $z \sim 1.5$ using the IR luminosity functions from \citet{Gruppionis13}.  The  magenta filled diamonds is the stepwise determination of the SFRF for $z \sim 0.9$ using the H$\alpha$ luminosity function from \citet{Ly11}. The dotted black line is the analytic SFRF that we obtained using the UV luminosity function from \citet[][ $z \sim1.9$]{Alavi13}. For the  grey squares we used the UV luminosity function from  \citet[][ $z \sim2$]{Parsa2015}. }
\label{fig:21SFRF}
\end{figure*}

To determine the SFRF of $z \sim2.0$ galaxies we use the IR luminosity function from \citet[][ $1.8 < z <2.3$]{Magenlli11} and the UV luminosity functions of \citet[][$z \sim2.0$]{Alavi13} and \citet[][ $z \sim2.0$]{Parsa2015}. We use equations \ref{eq:SFRpara2} and \ref{eq:SFRpara3} to convert the IR and UV luminosities to SFRs, respectively. The SFRF resulting from the IR data of \citet{Magenlli11} suggests that the UV selection of \citet{Alavi13} and \citet{Parsa2015} miss a significant number of objects at $z \sim2.0$ or that dust corrections implied by the IRX-$\beta$ relation and UV SFRs are underestimated at the bright-end of the distribution. This is similar with what we saw at redshift $z \sim2.6$ and section \ref{SFRF260}. The SFRF that is obtained from the data of \citet{Parsa2015} implies a distribution much shallower at low SFRs than that suggested by \citet{Magenlli11} and \citet{Alavi13}. The results from the above analysis are shown in Table \ref{tab_stepsfrf20} and Fig. \ref{fig:21SFRF}.

\begin{table}
  \centering
\resizebox{0.50\textwidth}{!}{%
  \begin{tabular}{cccc}
    \hline \\
    & {\large $\log\,\frac{{\rm SFR}}{{\rm M}_{\odot}\ {\rm yr}^{-1}}$}  &
    {\large $\log \, \phi_{\rm SFR}\ \left({\rm Mpc}^{-3}\ {\rm
          dex}^{-1}\right)$} \\ \\
    \hline \hline
    & $z\sim1.47$, \citet[][ H$\alpha$]{Sobral2013}  & 1 Mag, incompleteness checked \\ 
    \hline 
    & 1.00  & -2.13$\pm$0.10  \\
    & 1.10  & -2.25$\pm$0.09 \\
    & 1.20  & -2.34$\pm$0.06  \\
    & 1.30  & -2.47$\pm$0.05  \\
    & 1.40  & -2.62$\pm$0.05  \\
    & 1.50  & -2.73$\pm$0.04 \\
    & 1.60  & -2.91$\pm$0.08  \\
    & 1.70  & -3.18$\pm$0.11  \\
    & 1.80  & -3.55$\pm$0.18  \\
    & 1.90  & -3.81$\pm$0.26 \\
    & 2.00  & -4.22$\pm$0.38  \\
    & 2.10  & -4.55$\pm$0.55  \\
    & 2.30  & -4.86$\pm$0.55  \\
    \hline \hline
     & $z\sim1.3-1.8$, \citet[][ IR]{Magenlli11}  & No correction \\ 
    \hline 
    & 1.24  & -2.38$\pm$0.25  \\
    & 1.64  & -2.78$\pm$0.25 \\
    & 2.04  & -3.15$\pm^{0.25}_{0.26}$  \\
    & 2.44  & -3.69$\pm$0.26 \\
    & 2.84  & -4.75$\pm^{0.31}_{0.45}$ \\
    \hline \hline
     & $z\sim1.2-1.7$, \citet[][ IR]{Gruppionis13}  & No correction \\ 
    \hline 
    & 1.49  & -2.93$\pm$0.18  \\
    & 1.99  & -3.29$\pm$0.06 \\
    & 2.49  & -3.81$\pm$0.03  \\
    & 2.99  & -4.85$\pm$0.05 \\
    \hline \\
  \end{tabular}%
}
  \caption{Stepwise SFR functions at $z\sim1.5$ using the data from \citet{Sobral2013}, blue circles of Fig. \ref{fig:21SFRF} and \citet{Magenlli11}, red circles of Fig. \ref{fig:21SFRF}. The parameters of the analytic expression (dark green dot-dashed line) we obtain from dust correcting the \citet{schechter1976} fit given by \citet{cucciati12} are $\phi^{\star} = 0.0033$ ${\rm Mpc}^{-3}$, $SFR^{\star}=16.7$ $M_{\odot}/yr $ and $\alpha = -1.07$. }
  \label{tab_stepsfrf15}
\end{table}

\subsection{The Star formation rate function at $\boldsymbol{z \sim1.5}$}
\label{SFRF15}

To obtain the SFRF at redshift $z \sim1.5$ we use the UV luminosity function of \citet[][$ z\sim1.2-1.7$]{cucciati12} the IR luminosity functions from  \citet[][ $z\sim1.3-1.8$]{Magenlli11} and \citet[][ $z\sim 1.2-1.7$]{Gruppionis13} and the H$\alpha$ LF from \citet[][ $z \sim1.47$]{Sobral2013}.  The luminosity functions of \citet{Magenlli11} and \citet{Sobral2013} provide information about the intermediate and high star forming objects. The SFRFs from the IR luminosity functions imply that the dust corrections used to recover the H$\alpha$ luminosity function of \citet{Sobral2013} were underestimated for high star forming systems or that the H$\alpha$ selection of the authors  misses a significant number of dusty objects. As discussed in Subsection \ref{intr61} the 1 mag simplification that is commonly used for dust attenuation effects underestimates the intrinsic SFRs for $z > 0.3$. Maybe this is the reason of the inconsistency between the two SFRFs. The SFRF we obtain from the analytic UV luminosity function of \citet{cucciati12} provides information at low SFRs and implies a shallower distribution. The results are in perfect agreement with the SFRF obtained from \citet{Sobral2013} for high SFRs. Thus the IRX-$\beta$ relation may also underestimate the true SFR of high star forming objects. The characteristic SFR\footnote{The characteristic luminosity of a LF is the luminosity at which the power-law form becomes an exponential choke. We define the characteristic SFR as the SFR at which the behaviour of the SFRF changes from an exponential to a power-law.} implied by UV and H$\alpha$ data is much lower than that of IR studies. The SFRFs for $z \sim1.5$ from the above analysis are shown in the top right panel of Fig. \ref{fig:21SFRF} and Table \ref{tab_stepsfrf15}.

\subsection{The Star formation rate function at $\boldsymbol{z \sim1.15}$}
\label{SFRF115}

To obtain an estimate of the SFRF at redshift $z \sim1.15$ we use the IR luminosity functions from \citet[][$z \sim 1.0-1.2$]{Gruppionis13} and \citet[][$z \sim 1.0-1.3$]{Magnelli2013} and the FUV luminosity function of \citet[][ $z \sim1.0-1.2$]{cucciati12}.  Even at these intermediate redshifts, the IR surveys are unable to probe low star forming objects. The red dashed-line is the analytical SFRF obtained from the double power law fit of the LF of \citet{Magnelli2013}. The results from the FUV luminosity function of \citet{cucciati12} imply a much shallower SFRF with a much lower characteristic SFR. Once again we see that the SFRFs from the extrapolations of IR studies typically have higher characteristic SFRs and are steeper for faint objects than those found by UV data. Our results for the SFRF at redshift $z \sim1.15$ are in the bottom left panel of Fig.  \ref{fig:21SFRF} and Table \ref{tab_stepsfrf11}.

\begin{table}
  \centering
\resizebox{0.50\textwidth}{!}{%
  \begin{tabular}{cccc}
    \hline \\
    & {\large $\log\,\frac{{\rm SFR}}{{\rm M}_{\odot}\ {\rm yr}^{-1}}$}  &
    {\large $\log \, \phi_{\rm SFR}\ \left({\rm Mpc}^{-3}\ {\rm
          dex}^{-1}\right)$} \\ \\
    \hline \hline
    & $z\sim1.0-1.2$, \citet[][ IR selected]{Gruppionis13}  & No correction \\  
    \hline 
    & 1.49  & -2.80$\pm$0.09  \\
    & 1.99  & -3.17$\pm$0.06 \\
    & 2.49  & -4.00$\pm$0.03  \\
    & 2.99  & -5.18$\pm$0.12  \\
    \hline \hline
     & $z\sim1.0-1.3$, \citet[][ IR selected]{Magnelli2013}  & No correction  \\ 
    \hline 
    & 1.14  & -2.53$\pm$0.19  \\
    & 1.64  & -2.74$\pm^{0.17}_{0.18}$ \\
    & 1.84  & -2.98$\pm^{0.15}_{0.16}$  \\
    & 2.04  & -3.09$\pm^{0.19}_{0.22}$ \\
    & 2.24  & -3.58$\pm^{0.17}_{0.18}$ \\
    & 2.44  & -4.15$\pm^{0.19}_{4.15}$  \\
    & 2.64  & -4.47$\pm^{0.25}_{0.40}$ \\
    & 3.04  & -4.95$\pm^{0.34}_{4.95}$ \\
    \hline \\
  \end{tabular}%
}
  \caption{Stepwise SFR functions at $z\sim1.15$ using the data from \citet[][ red diamonds of Fig. \ref{fig:21SFRF}]{Gruppionis13} and \citet[][ red circles in the bottom left panel of Fig. \ref{fig:21SFRF}]{Magnelli2013}. The parameters of the analytic expression (dark green dot-dashed line) we obtain from dust correcting the \citet{schechter1976} fit given by \citet{cucciati12} are $\phi^{\star} = 0.006$ ${\rm Mpc}^{-3}$, $SFR^{\star}=8.3$ $M_{\odot}/yr $ and $\alpha = -0.93$. }
  \label{tab_stepsfrf11}
\end{table}

\begin{table}
  \centering
\resizebox{0.50\textwidth}{!}{%
  \begin{tabular}{cccc}
    \hline \\
    & {\large $\log\,\frac{{\rm SFR}}{{\rm M}_{\odot}\ {\rm yr}^{-1}}$}  &
    {\large $\log \, \phi_{\rm SFR}\ \left({\rm Mpc}^{-3}\ {\rm
          dex}^{-1}\right)$} \\ \\
    \hline \hline
     & $z\sim0.84$, \citet[][  H$\alpha$ selected]{Sobral2013}  & 1 mag, incompleteness checked \\ 
    \hline 
    & 0.60  & -1.93$\pm$0.03 \\
    & 0.75  & -2.02$\pm$0.03  \\
    & 0.90  & -2.18$\pm$0.04 \\
    & 1.05  & -2.43$\pm$0.06  \\
    & 1.20  & -2.73$\pm$0.17  \\
    & 1.35  & -3.01$\pm$0.17  \\
    & 1.50  & -3.27$\pm$0.21 \\
    & 1.65  & -3.79$\pm$0.55  \\
    & 1.80  & -4.13$\pm$1.51  \\
    \hline \hline
    & $z\sim0.8-1.0$, \citet[][ IR selected]{Magnelli2013} & No correction \\  
    \hline 
    & 1.49  & -3.09$\pm$0.08  \\
    & 1.99  & -3.24$\pm$0.04 \\
    & 2.49  & -4.23$\pm$0.05  \\
    & 2.99  & -5.74$\pm$0.25  \\
    \hline \hline
     & $z\sim0.7-1.0$, \citet[][ IR selected]{Gruppionis13}  & No correction  \\ 
    \hline 
    & 1.24  & -2.55$\pm^{0.13}_{0.15}$  \\
    & 1.44  & -2.59$\pm^{0.10}_{0.10}$ \\
    & 1.64  & -3.05$\pm^{0.16}_{0.20}$  \\
    & 1.84  & -3.14$\pm^{0.11}_{0.12}$ \\
    & 2.04  & -3.43$\pm^{0.20}_{0.32}$ \\
    & 2.24  & -3.88$\pm^{0.16}_{0.20}$  \\
    & 2.64  & -4.83$\pm^{0.31}_{4.83}$ \\
    \hline \hline
    & $z\sim0.84$, \citet[][ H$\alpha$ selected]{Ly11}  & dust corrected, incompleteness checked \\ 
    \hline 
    & -0.20  & -2.22$\pm$0.25  \\
    & -0.002  & -1.61$\pm$0.10 \\
    & 0.20  & -1.81$\pm$0.06  \\
    & 0.40  & -1.96$\pm$0.05  \\
    & 0.60  & -2.03$\pm$0.04  \\
    & 0.80  & -2.18$\pm$0.05 \\
    & 1.00  & -2.36$\pm$0.05  \\
    & 1.20  & -2.46$\pm$0.06  \\
    & 1.40  & -2.71$\pm$0.08  \\
    & 1.60  & -2.91$\pm$0.10 \\
    & 1.80  & -3.09$\pm$0.12  \\
    & 2.00  & -3.47$\pm$0.19  \\
    & 2.19  & -3.61$\pm$0.21  \\
    \hline \\
  \end{tabular}%
}
  \caption{Stepwise SFR functions at $z\sim0.9$ using the data from \citet{Ly11},  magenta diamonds of Fig. \ref{fig:21SFRF}, \citet{Gruppionis13}, red  diamonds in the bottom right of Fig. \ref{fig:21SFRF},  \citet{Magnelli2013}, red circles of Fig. \ref{fig:21SFRF} and \citet{Sobral2013},  blue circles of Fig. \ref{fig:21SFRF}. The parameters of the analytic expression (dark green dot-dashed line) we obtain from dust correcting the \citet{schechter1976} fit given by \citet{cucciati12} are $\phi^{\star} = 0.007$ ${\rm Mpc}^{-3}$, $SFR^{\star}=7.75$ $M_{\odot}/yr $ and $\alpha = -0.88$.}
\label{tab_stepsfrf09}
\end{table}

\subsection{The Star formation rate function at $\boldsymbol{z \sim0.9}$}
\label{SFRF090}

To retrieve the SFRF at redshift $z \sim0.9$ we use the luminosity functions of \citet[][ $z \sim 0.84$]{Ly11}, \citet[][ $\sim 0.8-1.0$]{Gruppionis13}, \citet[][ $\sim 0.8-1.0$]{cucciati12}, \citet[][ $z \sim 0.7-1.0$]{Magnelli2013} and \citet[][ $z \sim0.84$]{Sobral2013}. As discussed in Subsection \ref{intr61}, the luminosity function of \citet{Ly11} is an H$\alpha$ LF that is corrected for incompleteness and dust attenuation effects with the \citet{Hopkins01} dust correction law. This dust correction law implies a luminosity/SFR dependent correction with the observed (non intrinsic) luminosity/SFR. To obtain the SFRF from this H$\alpha$ luminosity function we use  equation \ref{eq:SFRpara1}. The magenta filled diamonds of Fig. \ref{fig:21SFRF} are the stepwise determination of the SFRF for $z \sim0.84$ from the H$\alpha$ luminosity function of \citet{Ly11}. From Fig. \ref{fig:21SFRF} we see that the bright-end of the SFRF from \citet{Ly11} is in excellent agreement with the results from the IR samples and this suggests that the SFR dependent dust corrections suggested by \citet{Hopkins01} are robust. For the low star forming objects we see that the SFRFs derived from \citet{Ly11} and \citet{Sobral2013} are in very good agreement. However, the SFRFs that rely on the results of \citet{Sobral2013} is highly uncertain for luminous star forming objects and is not consistent with the SFRs of the IR samples. \citet{Sobral2013} made a direct comparison between their H$\alpha$ luminosity function and that of \citet{Ly11} assuming the same dust correction law for both samples. The two luminosity functions are in excellent agreement. This indicates  that the 1 mag simplification is responsible for the tension with the IR SFRs and is not valid at $z \sim0.9$, since it underestimates the intrinsic SFRs/luminosities. Generally the 1mag correction is expected to overestimate the SFR for low mass (low dust contents) objects and underestimates it for high mass (high dust contents) objects, thus the SFRF from the results of \citet{Sobral2013} could be artificially steep. The results for the SFRF at $z \sim0.9$ from the above analysis are shown in the bottom right panel of Fig. \ref{fig:21SFRF} and Table \ref{tab_stepsfrf09}. \\ 
\\
In conclusion, SFRFs that rely on UV data are shallower than those obtained from IR LFs. The latter are unable to probe objects with low SFR and thus their extrapolation, which is commonly used to calculate the CSFRD, includes a lot of uncertainties. On the other hand, UV SFRFs-LFs are unable to succesfully probe high SFR systems due to the fact that they either fail to take into account dusty and massive objects or dust correction laws underestimate dust corrections, and hence intrinsic SFRs, in this range. Despite the differences at the faint and bright ends of the distribution, UV, H$\alpha$ and IR SFR indicators show excellent agreement for objects with $ -0.3 \leq \log(SFR/ (\, M_{\odot} \, yr^{-1})) \leq 1.5 $.

\begin{table*}
\centering
\resizebox{0.95\textwidth}{!}{%
\begin{tabular}{llccccccc}
  \\ \hline & Run & IMF & Box Size & N$_{\rm TOT}$ & M$_{\rm DM}$ & M$_{\rm GAS}$  &
  Comoving Softening & Feedback \\ & & & [Mpc/$h$] &
  & [M$_{\rm \odot}$/$h$] & [M$_{\rm \odot}$/$h$] & [kpc/$h$] \\ \hline
  & \textit{Kr24\tu eA\tu sW} & Kroupa & 24 & $2\times288^3$ &
  3.64$\times10^{7}$ & $7.32\times 10^6$ & 4.0 & Early AGN $+$ Constant strong Winds \\ \hline
  & \textit{Ch24\tu eA\tu nW} & Chabrier & 24 & $2\times288^3$ & 3.64$\times10^{7}$ & $7.32\times 10^6$ & 4.0 & Early AGN $+$ no Winds  \\ \hline
  & \textit{Ch24\tu NF} & Chabrier & 24 & $2\times288^3$ &
  3.64$\times10^{7}$ & $7.32\times 10^6$ & 4.0 & No Feedback\\ \hline
  & \textit{Ch24\tu eA\tu MDW}$^a$ & Chabrier & 24 & $2\times288^3$ &
  3.64$\times10^{7}$ & $7.32\times 10^6$ & 4.0 & Early AGN $+$ \\
  & & & & & & & & Momentum-Driven Winds \\ \hline
  & \textit{Ch24\tu eA\tu EDW}$^b$ & Chabrier & 24 & $2\times288^3$ &
  3.64$\times10^{7}$ & $7.32\times 10^6$ & 4.0 & Early AGN $+$ \\ 
  & & & & & & & & Energy-Driven Winds \\ \hline
\end{tabular}%
}
\caption{Summary of the different runs used in this work. Column 1, run name; column 2, Initial Mass Function (IMF) chosen; column 3, box size in comoving Mpc/$h$; column 4, total number of particles (N$_{\rm TOT} =$ N$_{\rm GAS}$ $+$ N$_{\rm DM}$); column 5, mass of the dark matter particles; column 6, initial mass of the gas particles; column 7, Plummer-equivalent comoving gravitational softening length; column 8, type of feedback implemented. See section \ref{thecode} and \citet{TescariKaW2013} for more details on the parameters used for the different feedback recipes. $(a)$: in this simulation we adopt variable momentum-driven galactic winds (Subection \ref{Feed}). $(b)$: in this simulation we adopt variable energy-driven galactic winds (Subsection \ref{Feed}).}
\label{tab:sim_runs}
\end{table*}

\section{Simulations}
\label{thecode}

In this work we use the set of {\textit{AustraliaN {\small{GADGET-3}} early Universe Simulations}} ({\textsc{Angus}}) described in \citet{TescariKaW2013}\footnote{The features of our code are extensively described in \citet{TescariKaW2013} and \citet{Katsianis2014}, therefore we refer the reader to those papers for additional information.}. We run these simulations using the hydrodynamic code {\small{P-GADGET3(XXL)}}. We assume a flat $\Lambda$ cold dark matter ($\Lambda$CDM) model with $\Omega_{\rm 0m}=0.272$, $\Omega_{\rm 0b}=0.0456$, $\Omega_{\rm \Lambda}=0.728$, $n_{\rm s}=0.963$, $H_{\rm 0}=70.4$ km s$^{-1}$ Mpc$^{-1}$ (i.e. $h = 0.704$) and $\sigma_{\rm 8}=0.809$. Our configurations have box size $L = 24$ Mpc/$h$, initial mass of the gas particles M$_{\rm GAS}=7.32\times 10^6$ M$_{\rm \odot}/h$ and a total number of particles equal to $2\times288^3$. All the simulations start at $z=60$ and were stopped at $z=0.8$. The different configurations were constrained at $z \sim4-7$ using the observations of \citet{smit12} in \citet{TescariKaW2013}. 

We explore different feedback prescriptions, in order to understand the origin of the difference between observed and simulated relationships.   We do not explore the broadest possible range of simulations, but concentrate on the simulations that can describe the high-$z$ star formation rate function and galaxy stellar mass function \citep{TescariKaW2013,Katsianis2014}. We performed resolution tests for high redshifts ($z \sim4-7$) in the appendix of \citet{Katsianis2014} and showed that our results converge for objects with $\log_{10} ({\rm M}{_\star}/{\rm M}_{\odot}) \ge 8.5$.

\subsection{SNe feedback}
\label{Feed}

We investigate the effect of three different galactic winds schemes in the simulated SFRF. First, we use the implementation of the constant galactic winds \citep{springel2003}. We assume the wind mass loading factor  $\eta=\dot{\rm M}_{\rm w}/\dot{\rm M}_{\rm \star}=2$ and a fixed wind velocity $v_{\rm w}=450$ km/s. \citet{PuchweinSpri12} demonstrated that constant wind models are not able to reproduce the observed galaxy stellar mass function at $z \sim 0$. Thus, similar to the authors, we explore as well the effects of variable wind models, in which the wind velocity is proportional to the escape velocity of the galaxy from which the wind is launched. This choice is supported by the observations of \citet{Martin05} who detect a positive correlation of galactic outflow speed with galaxy mass and showed that the outflow velocities are always $2$ to $3$ times larger than the galactic rotation speed. Inspired by these results, \citet{PuchweinSpri12} and \citet{barai13} assume that the velocity $v_{max}$ is related to the circular velocity $v_{circ}$. We use a momentum driven wind model in which the velocity of the winds is proportional to the circular velocity $v_{\rm circ}$ of the galaxy:
\begin{eqnarray}
  v_{\rm w}= 2\;\sqrt{\frac{G{\rm M}_{\rm halo}}{R_{\rm 200}}}=2\times v_{\rm circ},
\end{eqnarray}
and the loading factor $\eta$,
\begin{eqnarray}
  \eta = 2\times\frac{450\,\,{\rm km/s}}{v_{\rm w}},
\end{eqnarray}
where  M$_{\rm halo}$ is the halo mass and $R_{\rm 200}$ is the radius within which a density 200 times the mean density of the Universe at redshift $z$ is enclosed \citep{barai13}. Furthermore, we investigate the effect of the energy driven winds used by \citet{PuchweinSpri12}. In this case the loading factor is
\begin{eqnarray}
  \eta = 2\times\left(\frac{450\,\,{\rm km/s}}{v_{\rm w}}\right)^2,
\end{eqnarray}
while $v_{\rm w} = 2\times v_{\rm circ}$.

\subsection{AGN feedback}
\label{AGN}

In our scheme for Active Galactic Nuclei (AGN) feedback, when a dark matter halo reaches a mass above a given mass threshold  M$_{\rm th}=2.9\times10^{10}$ M$_{\rm\odot}/h$  for the first time, it is seeded with a central Super-Massive Black Hole (SMBH) of mass M$_{\rm seed}=5.8\times 10^{4}$  M$_{\rm\odot}/h$ (provided it contains a minimum mass fraction in stars $f_{\star}=2.0\times10^{-4}$). Each SMBH will then increase its mass through mergers or by accreting local gas from a maximum accretion radius $R_{\rm ac}=200$ kpc/$h$. In \citet{TescariKaW2013} we labeled the above feedback prescription as the early AGN feedback recipe. In this scheme we allow the presence of black hole seeds in relatively low mass halos (M$_{\rm th} \le 2.9\times10^{10}$ M$_{\rm\odot}/h$). Thus, SMBHs start to occupy dark matter halos at high redshifts and have enough time to grow and produce efficient feedback at $z \le 2$. The AGN feedback prescription that we use combined with efficient winds is successful at reproducing the observed SFRF \citep{TescariKaW2013} and GSMF \citep{Katsianis2014} for redshifts $ 4 < z < 7$.

\section{The Star formation rate function in hydrodynamic simulations}
\label{SFRFbhydro6}

\begin{figure*} 
\centering
\includegraphics[scale=0.83]{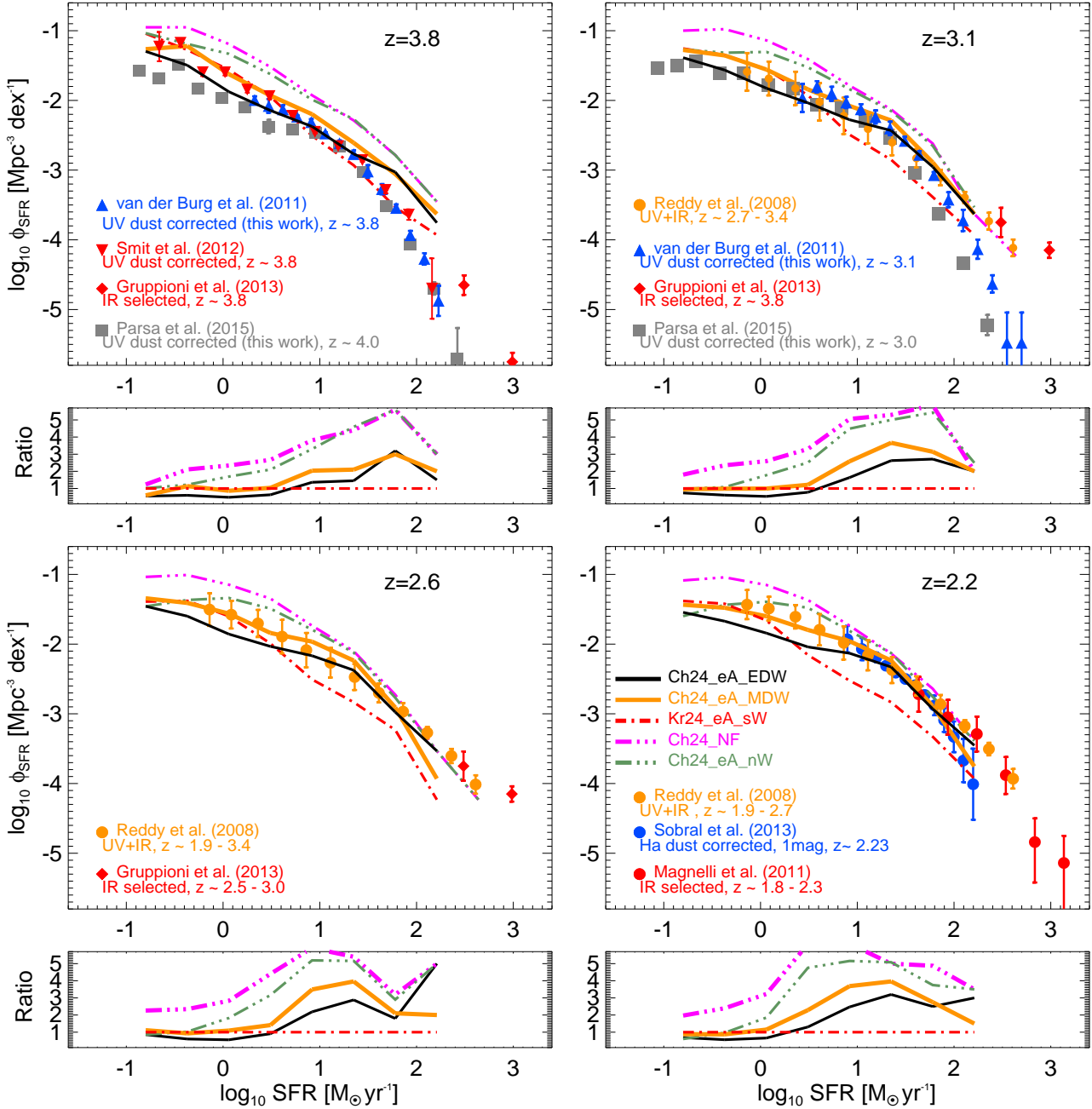}
\vspace{0.35cm}
\caption{The simulated SFRFs (lines) for redshifts $z \sim 3.8$ (top left panel),  $z \sim 3.1$ (top right panel), $z \sim 2.6$ (bottom left panel) and $z \sim 2.2$ (bottom right panel). Alongside we present the stepwise determinations of the observed SFRFs of Fig \ref{fig:38SFRF} for comparison.}
\label{SFRF42all}
\end{figure*}

In Fig. \ref{SFRF42all}  we present the evolution of the  star formation rate function from redshift $z \sim4.0$ to $z \sim2.2$ for our different runs and compare these with the observations discussed in Section \ref{Obsout6}. Like in \citet{TescariKaW2013} we name each run according to the IMF, boxsize and combination of feedback prescriptions that were used (more details can be found in table \ref{tab:sim_runs}). At each redshift, a panel showing ratios between the different simulations and the \textit{Kr24\tu eA\tu sW} run (red dot-dashed line) is included. The \textit{Kr24\tu eA\tu sW} run was the reference model used in \citet[][ SFRF, $z \sim4-7$]{TescariKaW2013} and \citet[][ GSMF, $z \sim4-7$]{Katsianis2014}. The constant wind model was used to model galactic winds in this case. This run produces similar results of simulations with variable galactic winds (Energy Driven Winds-EDW and Momentum Driven Winds-MDW)  at high redshift and we will use it as a reference in the following comparisons, despite the fact that it is not as successful at $z \sim1-4$. 

\begin{figure*}
\centering
\includegraphics[scale=0.83]{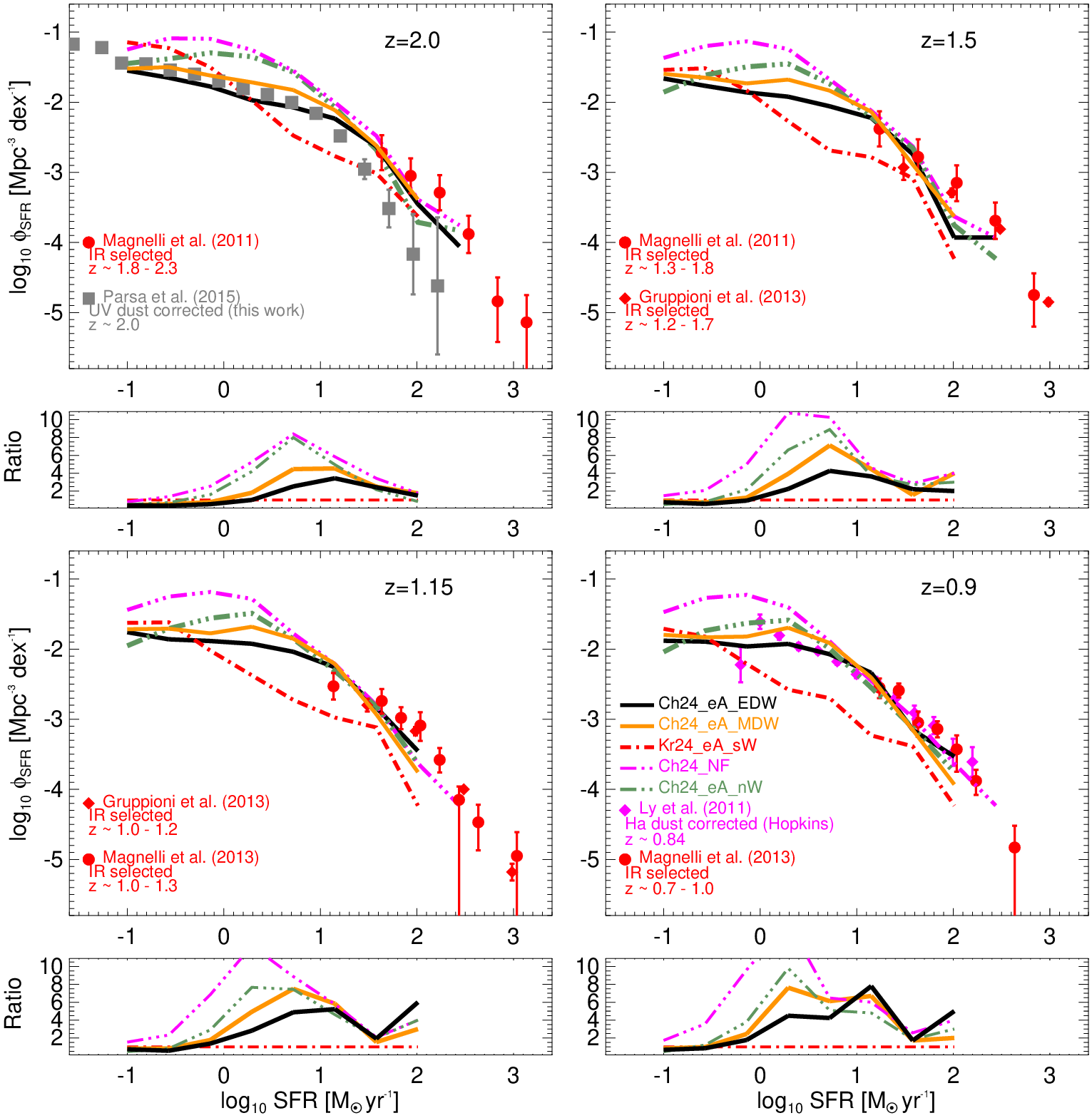}
\vspace{0.42cm}
\caption{The simulated SFRFs (lines) for redshifts $z \sim 2.0$ (top left panel), $z \sim 1.5$ (top right panel),  $z \sim 1.15$ (bottom left panel) and $z \sim 0.9$ (bottom right panel). We also present the stepwise determinations of the observed SFRFs of Fig.  \ref{fig:21SFRF} for comparison. }
\label{IntermediatelowSFRF}
\end{figure*}

At redshift $z = 3.8$ (top left panel of Fig. \ref{SFRF42all}) we see that the \textit{Ch24\tu NF} run (No feedback, magenta triple dot-dashed line), overproduces the number of systems with respect to all the other simulations and observations, due to the overcooling of gas. \citet{TescariKaW2013} discussed the effect of our feedback implementations on the simulated SFRF for redshifts $ 4 < z < 7 $ and suggested that some form of feedback is necessary. As discussed in \citet{TescariKaW2013}, the \textit{Kr24\tu eA\tu sW} and \textit{Ch24\tu eA\tu MDW} runs show a good consistency with observations from UV-selected samples \citep[e.g.][]{smit12}. Variations to the IMF  have a negligible impact on our simulated galaxy SFRF. The \textit{Ch24\tu eA\tu MDW} run slightly overproduces  high star forming objects since winds are less effective in this wind model for high mass/SFR systems. The  \textit{Ch24\tu eA\tu EDW} run  shows a good agreement with observations but underpredicts the number of objects with low star formation rates with respect to other runs. This is due to the fact that the winds of this model are very effective for low mass/SFR systems. We see that all the models implemented with galactic winds are able to broadly reproduce the observations, indicating that their presence is important.

We note a similar trend at redshift $z = 3.1$ (top right panel of Fig. \ref{SFRF42all}). The runs without winds overpredict the number of objects at all star formation rates. We see that the \textit{Kr24\tu eA\tu sW} run starts to underpredict objects with $\log(SFR/ (\, M_{\odot} \, yr^{-1})) \geq 0.5$ with respect the \textit{Ch24\tu eA\tu MDW} and \textit{Ch24\tu eA\tu EDW} runs. The last two have better consistency with observations. The \textit{Ch24\tu eA\tu MDW} and \textit{Ch24\tu eA\tu EDW} runs produce almost identical SFRFs for objects with $\log(SFR/ (\, M_{\odot} \, yr^{-1})) \geq 0.5$, but the second produces less objects with $\log(SFR/ (\, M_{\odot} \, yr^{-1})) \leq 0.5$ due to the fact that  energy variable driven winds are more effective in this range. This brings simulations into better agreement with the data of \citet{Parsa2015}.

At redshift $z = 2.6$ (bottom left panel of Fig. \ref{SFRF42all}) we see once again that the simulation with constant energy-driven winds tend to underpredict objects with high star formation rates. There is no need to strongly quench the SFR of high star forming objects and all the configurations, including the \textit{Ch24\tu NF} run, are consistent with the observations for objects with $\log(SFR/ (\, M_{\odot} \, yr^{-1})) \geq 1.5$. It is necessary though to have a feedback prescription to decrease the SFRs of objects with  $\log(SFR/ (\, M_{\odot} \, yr^{-1})) \leq 1.5$. The efficient variable momentum and energy driven winds are good candidates. The \textit{Ch24\tu eA\tu MDW} and \textit{Ch24\tu eA\tu EDW} runs are consistent with the observations, even though the \textit{Ch24\tu eA\tu EDW} run slightly underpredicts the number of objects with low star formation rates.

At redshift $z = 2.2$ (bottom right panel of Fig. \ref{SFRF42all}) we can see the effect of different feedback prescriptions more clearly. This era represents the peak of the cosmic star formation rate density, and so it is anticipated that feedback related to stars and SNe will play an important role in the regulation of star formation. Constant winds are very efficient for objects with high star formation rates and these are common at this epoch. Interestingly, we note that all simulations except \textit{Kr24\tu eA\tu sW} and \textit{Ch24\tu NF} are broadly consistent with observations at $z = 2.2$, but there is a requirement to decrease the number of objects with low star formation rates.  The \textit{Ch24\tu eA\tu nW} and \textit{Ch24\tu eA\tu MDW} runs are consistent with the observations, even though the \textit{Ch24\tu eA\tu EDW} run slightly underpredicts the number of objects with $\log(SFR/ (\, M_{\odot} \, yr^{-1})) \leq 0.3$. Despite this, in the following paragraphs we will see that this run is the most successful at lower redshifts because it is able to reproduce the shallow SFRFs obtained from UV LFs.

At redshift $z = 2.0$ (top left panel of Fig. \ref{IntermediatelowSFRF}) we  find that simulations with variable winds are quite successful at reproducing the SFRF implied by the UV luminosity function of \citet{Parsa2015}. The \textit{Kr24\tu eA\tu sW}  run underpredicts objects with high star formation rate ($\log(SFR/ (\, M_{\odot} \, yr^{-1})) \geq 0.5$), with respect to all other runs. The run without feedback \textit{Ch24\tu NF} overpredicts the number of objects with low star formation rates, but has good agreement with the constraints from IR studies. This indicates that at $z \sim2.0$  there is no need for feedback to regulate the SFR of high SFR objects in our simulations. However, efficient feedback is necessary to decrease the number of objects with low star formation rate, and variable enegy driven winds are perfect candidates.

\begin{figure*}
\centering
\includegraphics[scale=0.83]{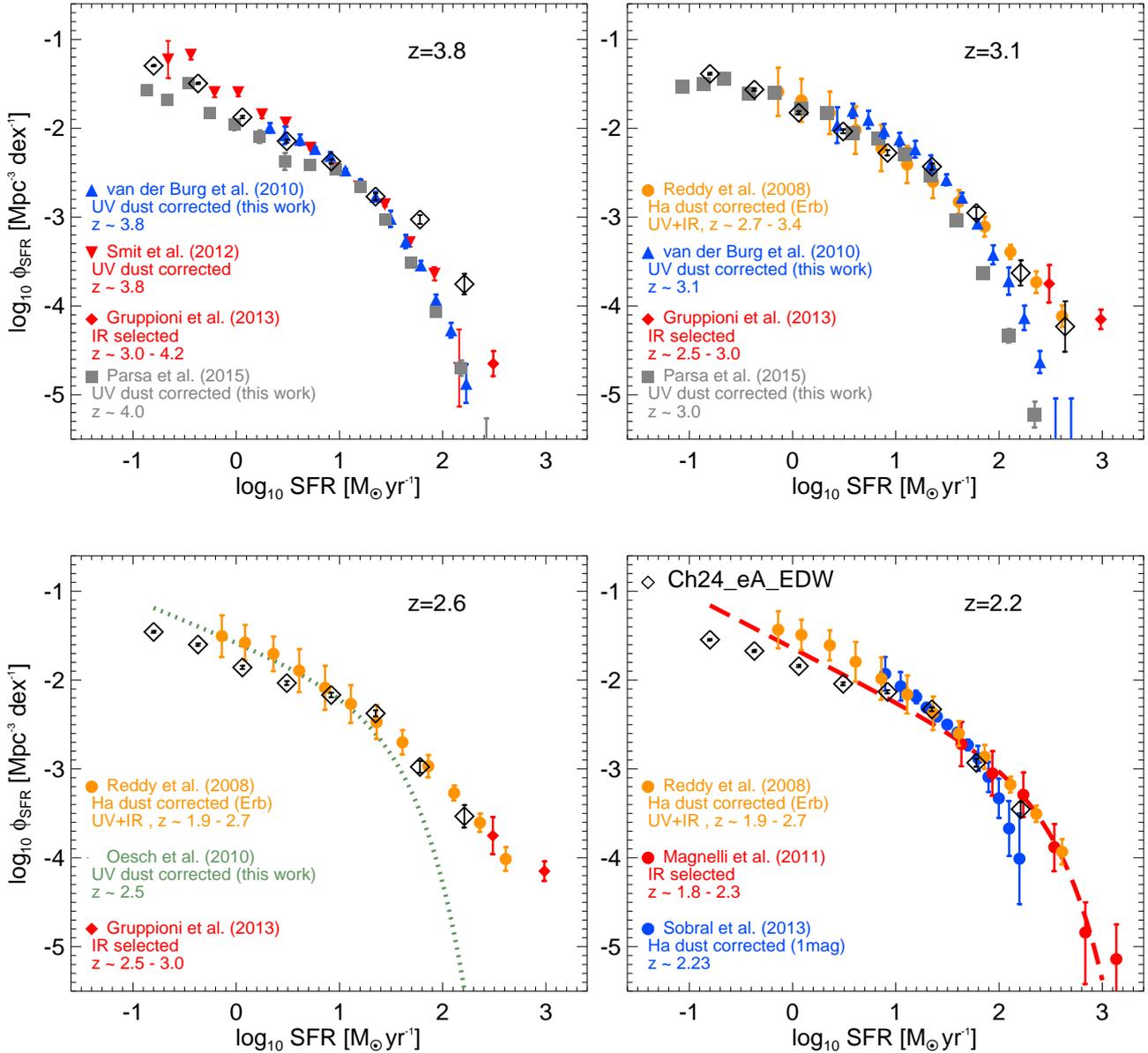}
\vspace{0.42cm}
\caption{The stepwise and analytical determinations of the observed SFRF along with our fiducial model for redshifts $z \sim 3.8$ (top left panel), $z \sim 3.1$ (top right panel),  $z \sim 2.6$ (bottom left panel) and  $z \sim 2.2$ (bottom right panel).}
\label{HighIntermediateSFRFid}
\end{figure*}

\begin{figure*} 
\centering
\includegraphics[scale=0.83]{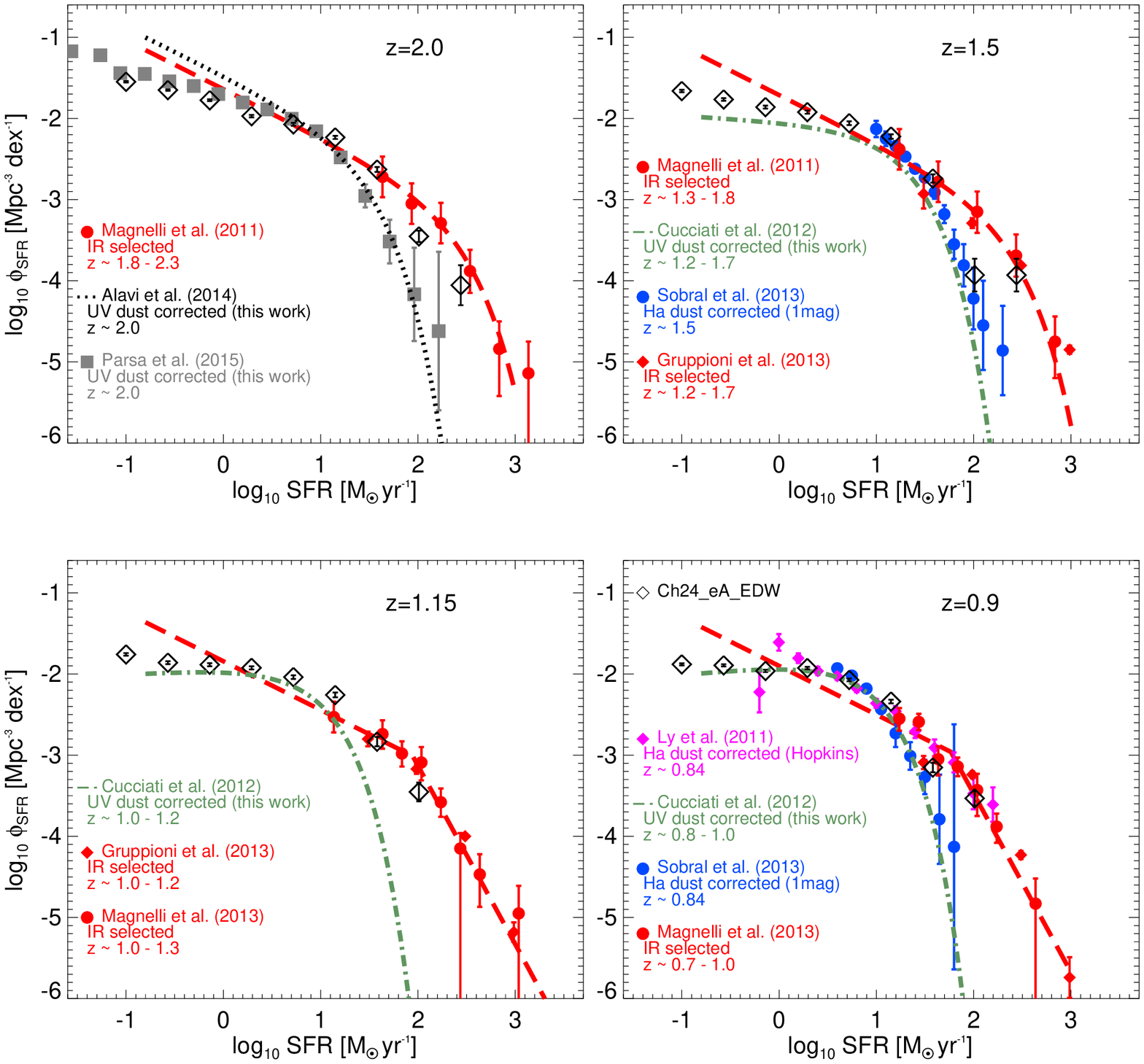}
\vspace{0.42cm}
\caption{The stepwise and analytical determinations of the SFRF alongside with our best fiducial model for redshifts  $z \sim 2.0$ (top left panel), $z \sim 1.5$ (top right panel), $z \sim 1.15$ (bottom left panel) and $z \sim 0.9$ (bottom right panel). }
\label{IntermediatelowSFRFid}
\end{figure*}

At redshift $z = 1.5$ (top right panel of Fig. \ref{IntermediatelowSFRF}) and $z = 1.15$ (bottom left panel of Fig. \ref{IntermediatelowSFRF}) we see that feedback prescriptions with variable galactic winds are quite succesfull at reproducing the SFRF implied by the IR luminosity functions, while the \textit{Kr24\tu eA\tu sW}  run underpredicts objects with high and intermediate star formation. On the other hand, the \textit{Ch24\tu eA\tu EDW} and \textit{Ch24\tu eA\tu MDW} runs are once again able to reproduce the observations. This is also true for redshift $z = 0.9$ (bottom right panel of Fig. \ref{IntermediatelowSFRF}). We see that the run without feedback, \textit{Ch24\tu NF}, overpredicts the number of objects with low star formation rate but the difference with observations and the rest of the runs is much smaller than at higher redshifts. This could imply the following. \\ A) It is possible that galaxies in the \textit{Ch24\tu NF} run  depleted their gas at high redshifts, where the SFRs of the objects were very high at early times. The SFRF in the no feedback scenario at $z \sim4$ is almost 7 times larger than the observations. This could explain the small difference between the SFRFs of the \textit{Ch24\tu NF} and  \textit{Ch24\tu eA\tu EDW} configurations at low redshifts, especially at high star forming objects. \\ B) To reproduce the observed evolution of the SFRF we need efficient feedback at early times, while at lower redshifts we require the scheme to become relatively moderate. The \textit{Ch24\tu eA\tu EDW} and \textit{Ch24\tu eA\tu MDW} runs are successful at reproducing the observations. We saw in section \ref{thecode} that variable galactic winds are successful at decreasing the SFRs of objects which reside low mass halos.  Galaxies reside typically in low mass in the early Universe, so overall this feedback prescription is quite efficient at high redshifts. On the contrary, the scheme becomes relatively moderate at decreasing the SFRs of objects  at lower redshifts where halos have become larger.

In conclusion, the simulation that does not take into account any form of feedback, is consistent with the observed SFRF at low and intermediate redshifts, despite the fact that it is in tension with observables at $z > 2.0$. This is a strong indication that the efficiency of feedback prescriptions in our simulations should decrease with time. By construction variable galactic winds are efficient at decreasing the SFR of objects with low mass. Overall this prescription should be more efficient at high redshifts where haloes are typically smaller. On the other hand, at low redshifts halos are typically larger and this results in winds that become less efficient. This behaviour makes the variable winds a good choice to model galactic outflows in our simulations.

\subsection{Best fiducial model}

\begin{figure*} 
\centering
\includegraphics[scale=0.83]{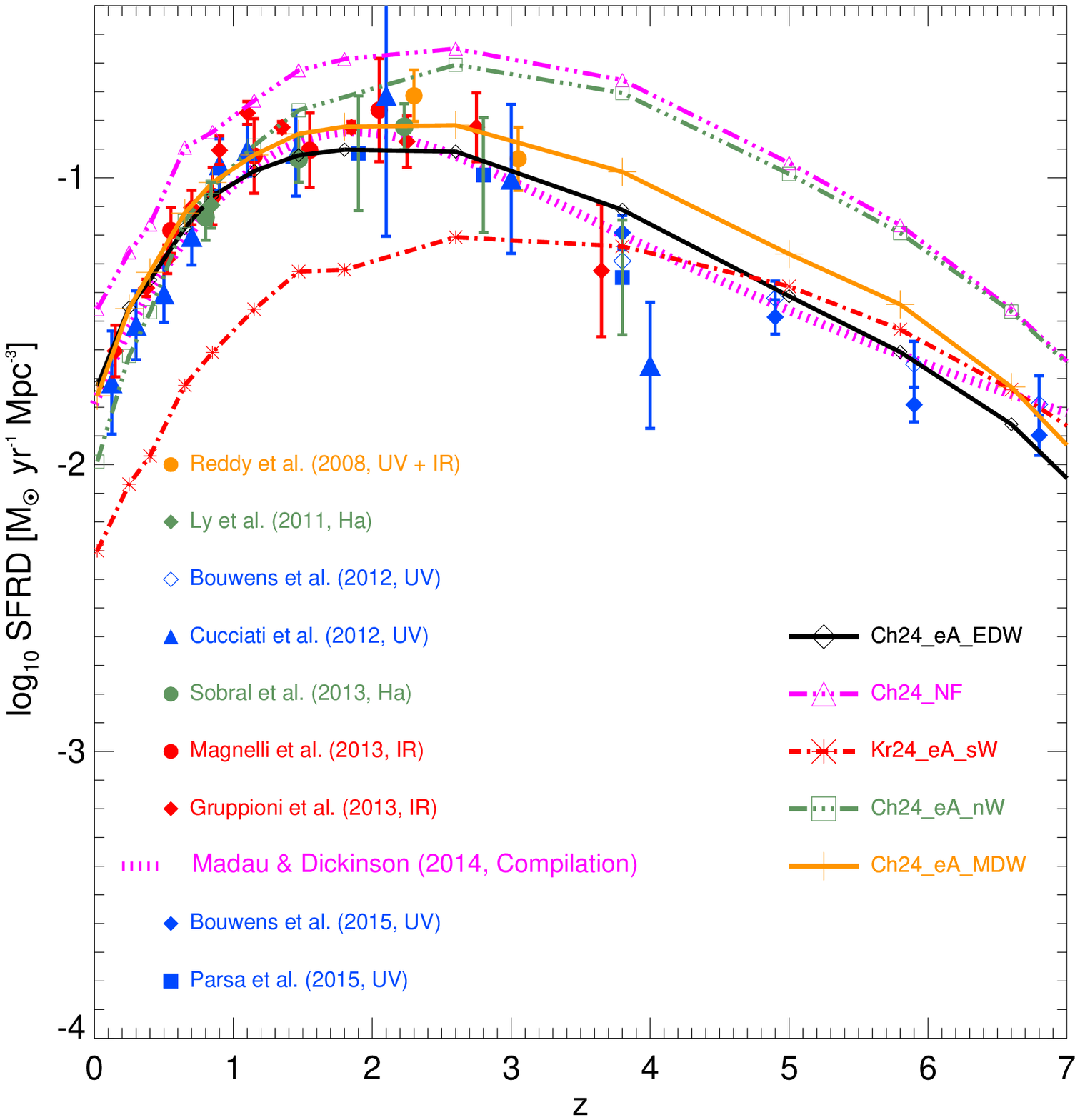}
\caption{The evolution of the cosmic star formation rate density from $z \sim7$ to $z \sim0$ in cosmological hydrodynamic simulations and observations. The orange filled circles represent the results of \citet{Reddy2008} .  The red filled reversed triangles, circles and diamonds represent the CSFRD from the integration of the SFRFs implied by the IR luminosity functions of   \citet{Magnelli2013} and \citet{Gruppionis13}, respectively. The dark green filled diamonds and green filled circles use the  H$\alpha$ data of \citet{Ly11} and \citet{Sobral2013}, respectively. The blue filled triangles, open diamonds and filled squares represent the CSFRD obtained from UV luminosity functions of \citet{cucciati12}, \citet{bouwens2012}, \citet{Bouwens2014} and \citet{Parsa2015}, respectively.  The black dotted line represents the compilation study of \citet{Madau2014}. The luminosity limit is set to $L_{min} = 0.03 \, L^{\star}$ as in \citet{Madau2014}. }
\label{CSFRDall}
\end{figure*}

In Figs. \ref{HighIntermediateSFRFid} and \ref{IntermediatelowSFRFid} we see that the configuration that has the best agreement with observations for all the redshifts considered in this work is the \textit{Ch24\tu eA\tu EDW} run, which combines a Chabrier IMF, early AGN feedback and energy-driven winds. In Fig. \ref{HighIntermediateSFRFid} we show SFRFs at redshift $z\sim2.2-4$ for our fiducial model (open black diamonds with error bars), alongside the stepwise and analytical determinations of the observed SFRF already presented in Fig. \ref{fig:38SFRF}.  We include Poissonian uncertainties for the simulated SFRFs (black error bars), in order to provide an estimate of the errors from our finite box size. We see that this model is able to reproduce the SFRFs derived from the IR, UV and H$\alpha$ studies. For low  luminosity/star forming objects the fiducial run is able to obtain the shallow SFRFs of faint objects implied by UV data, while also being in agreement with the constraints from IR studies for high luminosity/star forming systems. Moving to lower redshifts (Fig. \ref{IntermediatelowSFRFid}), we can see a great consistency of the fiducial model with the observations presented in Fig. \ref{fig:21SFRF}. The simulated SFRFs are in good agreement with the UV and H$\alpha$ studies for objects with  $ -1.0 \leq \log(SFR/ (\, M_{\odot} \, yr^{-1})) \leq 1.0 $, and with IR data for $\log(SFR/ (\, M_{\odot} \, yr^{-1})) \geq 1.0$.  The variable energy driven winds that efficiently decrease the SFR of low mass objects are quite successful at reproducing the shallow SFRFs implied by UV data, and also have good agreement with the constraints from IR studies for the high star forming and dusty objects.

\section{The simulated and observed cosmic star formation rate density}
\label{CSFRDsim}

The evolution of the cosmic star formation rate density of the Universe is commonly used to test theoretical models, since it represents a fundamental constraint on the growth of stellar mass in galaxies over time. In the above Sections we saw that the SFRFs can commonly be described by the \citet{schechter1976} functional form. The integration of the \citet{schechter1976} fit gives the total Cosmic Star Formation Rate Density (CSFRD) of the Universe at a given redshift. It is usual for observers to set limits on to the integration of the LFs when calculating the cosmic luminosity density (LD) that is then converted to CSFRD. Usually, the assumed lower cut corresponds to the sensitivity of the observations available. \citet{Madau2014} used a compilation of luminosity functions to constrain the evolution of CSFRD. The integration limit that the authors chose was  $L_{min} = 0.03 \, L^{\star}$, where $L^{\star}$ is the characteristic luminosity. The above integration can be written as
\begin{eqnarray}
  p_{\rm LD}= \int_{0.03 \, L^{\star}}^{\infty} L \,  \phi(L,z) \, dL,
\end{eqnarray}
and the integration gives 
\begin{eqnarray}
\label{laaa}
  p_{\rm LD} = \Gamma (2 \, + \, \alpha, \, 0.03) \, \phi^{\star} \, L^{\star},
\end{eqnarray}
where $\alpha$ is the faint-end slope, which describes how steep the LF is. From equation \ref{laaa} we see that higher values of characteristic luminosity $L^{\star}$ and faint-end slope $\alpha$ result in higher values of luminosity density (i.e. shallower LFs with small characteristic luminosities give lower LD). After the cosmic LD is calculated, the \citet{kennicutt1998} relation can be employed to convert the luminosity density to SFR density. We use the same  method and lower limits to integrate the SFRFs presented in Section \ref{Obsout6}. We also include the  $z \sim4-8$ observations of \citet{bouwens2012,Bouwens2014} and the $z \sim0-0.7$ observations of \citet{cucciati12}, \citet{Magnelli2013} and \citet{Gruppionis13}. We present the results in Fig. \ref{CSFRDall}. The results originating from UV observations are in blue, the H$\alpha$ in green and the IR in red. \citet{Reddy2008} CSFRD originate from a bolometric (UV+IR) luminosity and is in orange. The black dotted line represents the compilation study of \citet{Madau2014}.

In the previous Sections we saw that the majority of the SFRFs that we obtained from dust corrected UV and H$\alpha$ LFs are shallower than those from the IR LFs for faint objects. The faint-end slope of the IR SFRF and LF is not directly constrained by individually detected sources and relies on extrapolations. In addition, we demonstrated that IR LFs can more successfully probe dusty, high star forming objects and therefore the SFRFs they produce, have higher characteristic SFRs than UV and H$\alpha$ data. These two factors can lead to overestimations in the calculations of the CSFRD that rely solely on IR data. 

In agreement with previous work in the literature \citep{Madau2014} we find that different SFR indicators produce consistent results for the CSFRD and this occurs despite differences in the measurements at the faint and bright-end of the SFRF. This is most likely due to the fact that all SFR tracers agree quite well for objects with $ -0.3 \leq \log(SFR/ (\, M_{\odot} \, yr^{-1})) \leq 1.5 $ (close to the characteristic SFR), which dominate the CSFRD at $z \sim1-4$. However, the results from IR SFRFs and LFs are typically $0.10-0.25$ dex larger. We find that systematics between different SFR indicators do not significantly affect the measurements of the CSFRD despite their differences for low (IR) and high star forming systems (UV, H$\alpha$).

In figure \ref{CSFRDall} we present as well the evolution of the simulated CSFRD for our simulations alongside with the observational constraints. The lower limit of the SFR cut for the simulations has been chosen to match the lower limit assumed in the observations. The simulation with no feedback (\textit{Ch24\tu NF}) overproduces the CSFRD at all redshifts considered and the peak of star formation activity is at $z \sim2.5$. At $z \sim1-2.5$ the CSFRD decreases slowly with time, while at $z \sim0-1$ the decrement is much faster. This maybe suggests that gas reservoirs were consumed at $z \sim 2$, when the SFR was high and no gas was left to fuel star formation at lower redshifts. If we take into account AGN feedback (\textit{Ch24\tu eA\tu nW}) we can see that the simulated CSFRD has a peak once again at $z \sim2.5$, but starts to decrease and becomes consistent with the observations for $z \sim0-1$. However, at $z \geq 1.5$ feedback from supermassive black holes is not sufficient to bring observations and simulations in agreement due to the fact that they are not large enough to produce the required energy to quench the star formation. The presence of another feedback mechanism is required to decrease the simulated CSFRD at $z \geq 1.5$. The \textit{Ch24\tu eA\tu EDW} and \textit{Ch24\tu eA\tu MDW} runs are implemented with feedback prescriptions that are efficient at high redshifts, making them good candidates. The \textit{Ch24\tu eA\tu EDW} run that successfully reproduced the observed SFRF also has an excellent agreement with the constraints from the observed CSFRD. The star formation peak occurs later than in the case with no feedback ($z \sim2.0$) and has a value $0.3$ dex lower. The run with constant energy galactic winds has good agreement with observations at $z \sim4-7$. However, at $z < 3.5$ it underpredicts the CSFRD with respect to the rest of the simulations. In Section \ref{SFRFbhydro6}, we demonstrated that this feedback prescription is very efficient for objects with high SFRs. In Fig. \ref{CSFRDall} we see that constant galactic winds are very efficient at low redshifts and decrease the CSFRD substantially. The tension with observations becomes more severe with time (0.3 dex at $z \sim3.0$, 0.5 dex at $z \sim1.0$ and 0.7 dex at $z \sim0$). 

In conclusion, the early AGN feedback prescription employed in our model decreases the cosmic star formation rate density at $z \leq 3$ but is not sufficient to reproduce the observed evolution of the cosmic star formation rate density at high redshifts ($z \geq 1.5$), since SMBHs are not massive enough to release sufficient energy. Variable galactic winds are perfect candidates to reproduce the observables, since their efficiency is large at high redshifts, but decreases at lower redshifts.

\section{Conclusions}
\label{concl6}

In this paper, we  investigated the evolution of the galaxy star formation rate function (SFRF) at $z \sim1-4$. In particular, we have focused on the role of supernova driven galactic wind and AGN feedback. We explored the effects of implementations of SN driven galactic winds presented in \citet{springel2003}, \citet{PuchweinSpri12} and \citet{TescariKaW2013}. For the first case, we explored a wind configuration (constant velocity $v_{\rm w}=450$). We also adopted variable momentum-driven galactic winds following \citet{TescariKaW2013} and variable energy-driven galactic winds following \citet{PuchweinSpri12}.

In the following we summarise the main results and conclusions of our analysis:
\begin{itemize}

\item The comparison between the SFRFs from H$\alpha$ and IR luminosities favour a luminosity/SFR dependent dust correction to the observed H$\alpha$ luminosities/SFRs. This is in agreement with other authors in the literature \citep[e.g.,][]{Hopkins01,cucciati12}. The IR luminosities provide a good test of dust physics and an appropriate indicator for the intrinsic SFRs. However, these rely on uncertain extrapolation to probe the SFR of low luminosity star forming objects. The H$\alpha$ and UV luminosity functions that are corrected for dust attenuation effects produce SFRFs that are consistent with IR data at intermediate SFRs. H$\alpha$ and UV SFRF-LFs are able to probe low mass objects, unlike IR derivations of the SFRF, but are either unable to probe the full population of high star formation rate objects or the dust correction implied by the IRX-$\beta$ relation underestimates the amount of dust. This suggests that IR and UV data have to be combined to correctly probe the SFRs of galaxies at both the faint and bright ends of the distribution and that systematic between SFR indicators can affect the measurements of the SFRF. The SFRFs that rely on UV data are shallower than those obtained from IR LFs, with lower characteristic SFRs. Despite their differences at the faint and bright-ends of the distribution, UV, H$\alpha$ and IR SFR indicators are in excellent agreement for objects with $ -0.3 \leq \log(SFR/ (\, M_{\odot} \, yr^{-1})) \leq 1.5$.   

\item Different SFR indicators produce consistent results for the CSFRD, despite their differences in the faint and bright-end of the SFRF. This is most likely due to the fact that all SFR tracers agree well for objects with $ -0.3 \leq \log(SFR/ (\, M_{\odot} \, yr^{-1})) \leq 1.5$, which dominate the CSFRD at $z \sim1-4$. However, the results from IR SFRFs and LFs are typically $0.10-0.25$ dex larger. This is due to the fact that the faint-end slopes of the IR SFRF and LF are not directly constrained by individually detected sources and rely only on extrapolations which have artificially smaller negative slope $\alpha$. In addition, the characteristic luminosity-SFR of IR studies is higher than that from UV and H$\alpha$ studies, which are unable to trace dusty systems with high star formation rates. Overall, systematics between different SFR indicators do not significantly affect the measurements of the CSFRD,  despite the inaccuracies for low (IR) and high star forming systems (UV, H$\alpha$).

\item The simulation that does not take into account any form of feedback, is consistent with the observed SFRF at the low and intermediate redshifts considered in this work (especially for high star forming objects), despite the fact that it is in disagreement with observables at $z > 2.0$. This is a strong indication that in our simulations the efficiency of feedback prescriptions should decrease with time. By construction, variable galactic winds are efficient at decreasing the SFR of objects with low mass. Overall, this prescription should therefore be more efficient at high redshifts where halos are typically smaller. On the other hand, at low redshifts halos are typically larger because they had the time to grow through accretion, which results in winds that become less efficient. This behaviour makes variable winds a good choice to model galactic outflows in our simulations.

\item The early AGN feedback prescription that we use decreases the cosmic star formation rate density at $z \leq 3$, but is not sufficient to reproduce the observed evolution of the cosmic star formation rate density at $z > 3$, since SMBHs are not massive enough to release enough energy at $z \geq 1.5$. On the other hand, variable galactic winds are perfect candidates to reproduce the observables since their efficiency is large at high redshifts and decreases at lower redshifts. The \textit{Ch24\tu eA\tu EDW} run  is the simulation that performs best overall.

\end{itemize}

In conclusion, we favour galactic winds that produce feedback that becomes less efficient with time. We favour feedback prescriptions that decrease the number of objects with low star formation rates.

\section*{Acknowledgments}

We would like to thank Volker Springel for making available to us the non-public version of the {\small{GADGET-3}} code. We would also like to thank Stuart Wyithe, Kristian Finlator, Lee Spitler and the anonymous referee for their comments. This research was conducted by the Australian Research Council Centre of Excellence for All-sky Astrophysics (CAASTRO), through project number CE110001020. This work was supported by the NCI National Facility at the ANU,  the Melbourne International Research Scholarship (MIRS) scholarship and the Albert Shimmins Fund - writing-up award provided by the University of Melbourne.  AK is supported by the CONICYT-FONDECYT fellowship (project number: 3160049).

\bibliographystyle{mn2e}	
\bibliography{Katsianis_mnrasRev}

\newcommand{\noopsort}[1]{}
\begin{thebibliography}{}
\makeatletter
\relax
\def\mn@urlcharsother{\let\do\@makeother \do\$\do\&\do\#\do\^\do\_\do\%\do\~}
\def\mn@doi{\begingroup\mn@urlcharsother \@ifnextchar [ {\mn@doi@}
  {\mn@doi@[]}}
\def\mn@doi@[#1]#2{\def\@tempa{#1}\ifx\@tempa\@empty \href
  {http://dx.doi.org/#2} {doi:#2}\else \href {http://dx.doi.org/#2} {#1}\fi
  \endgroup}
\def\mn@eprint#1#2{\mn@eprint@#1:#2::\@nil}
\def\mn@eprint@arXiv#1{\href {http://arxiv.org/abs/#1} {{\tt arXiv:#1}}}
\def\mn@eprint@dblp#1{\href {http://dblp.uni-trier.de/rec/bibtex/#1.xml}
  {dblp:#1}}
\def\mn@eprint@#1:#2:#3:#4\@nil{\def\@tempa {#1}\def\@tempb {#2}\def\@tempc
  {#3}\ifx \@tempc \@empty \let \@tempc \@tempb \let \@tempb \@tempa \fi \ifx
  \@tempb \@empty \def\@tempb {arXiv}\fi \@ifundefined
  {mn@eprint@\@tempb}{\@tempb:\@tempc}{\expandafter \expandafter \csname
  mn@eprint@\@tempb\endcsname \expandafter{\@tempc}}}

\bibitem[\protect\citeauthoryear{{Alavi} et~al.,}{{Alavi}
  et~al.}{2014}]{Alavi13}
{Alavi} A.,  et~al., 2014, \mn@doi [\apj] {10.1088/0004-637X/780/2/143}, \href
  {http://adsabs.harvard.edu/abs/2014ApJ...780..143A} {780, 143}

\bibitem[\protect\citeauthoryear{{Barai} et~al.,}{{Barai}
  et~al.}{2013}]{barai13}
{Barai} P.,  et~al., 2013, \mn@doi [\mnras] {10.1093/mnras/stt125}, \href
  {http://adsabs.harvard.edu/abs/2013MNRAS.430.3213B} {430, 3213}

\bibitem[\protect\citeauthoryear{{Bell}, {Zheng}, {Papovich}, {Borch}, {Wolf}
  \& {Meisenheimer}}{{Bell} et~al.}{2007}]{Bell07}
{Bell} E.~F.,  {Zheng} X.~Z.,  {Papovich} C.,  {Borch} A.,  {Wolf} C.,
  {Meisenheimer} K.,  2007, \mn@doi [\apj] {10.1086/518594}, \href
  {http://adsabs.harvard.edu/abs/2007ApJ...663..834B} {663, 834}

\bibitem[\protect\citeauthoryear{{Bouwens}, {Illingworth}, {Franx}  \&
  {Ford}}{{Bouwens} et~al.}{2007}]{bouwens2007}
{Bouwens} R.~J.,  {Illingworth} G.~D.,  {Franx} M.,   {Ford} H.,  2007, \mn@doi
  [\apj] {10.1086/521811}, \href
  {http://adsabs.harvard.edu/abs/2007ApJ...670..928B} {670, 928}

\bibitem[\protect\citeauthoryear{{Bouwens} et~al.,}{{Bouwens}
  et~al.}{2009}]{bouwens09}
{Bouwens} R.~J.,  et~al., 2009, \mn@doi [\apj] {10.1088/0004-637X/705/1/936},
  \href {http://adsabs.harvard.edu/abs/2009ApJ...705..936B} {705, 936}

\bibitem[\protect\citeauthoryear{{Bouwens} et~al.,}{{Bouwens}
  et~al.}{2012}]{bouwens2012}
{Bouwens} R.~J.,  et~al., 2012, \mn@doi [\apj] {10.1088/0004-637X/754/2/83},
  \href {http://adsabs.harvard.edu/abs/2012ApJ...754...83B} {754, 83}

\bibitem[\protect\citeauthoryear{{Bouwens} et~al.,}{{Bouwens}
  et~al.}{2015}]{Bouwens2014}
{Bouwens} R.~J.,  et~al., 2015, \mn@doi [\apj] {10.1088/0004-637X/803/1/34},
  \href {http://adsabs.harvard.edu/abs/2015ApJ...803...34B} {803, 34}

\bibitem[\protect\citeauthoryear{{Calzetti}, {Kennicutt}, {Engelbracht},
  {Leitherer}, {Draine}, {Kewley}, {Moustakas}  et~al.}{{Calzetti}
  et~al.}{2007}]{Calzetti07}
{Calzetti} D.,  {Kennicutt} R.~C.,  {Engelbracht} C.~W.,  {Leitherer} C.,
  {Draine} B.~T.,  {Kewley} L.,  {Moustakas} J.,   et~al., 2007, \mn@doi [\apj]
  {10.1086/520082}, \href {http://adsabs.harvard.edu/abs/2007ApJ...666..870C}
  {666, 870}

\bibitem[\protect\citeauthoryear{{Cucciati} et~al.}{{Cucciati}
  et~al.}{2012}]{cucciati12}
{Cucciati} O.,  et~al., 2012, \mn@doi [\aap] {10.1051/0004-6361/201118010},
  \href {http://adsabs.harvard.edu/abs/2012A%26A...539A..31C} {539, A31}

\bibitem[\protect\citeauthoryear{{Dav{\'e}}, {Oppenheimer}  \&
  {Finlator}}{{Dav{\'e}} et~al.}{2011}]{Dave2011}
{Dav{\'e}} R.,  {Oppenheimer} B.~D.,   {Finlator} K.,  2011, \mn@doi [\mnras]
  {10.1111/j.1365-2966.2011.18680.x}, \href
  {http://adsabs.harvard.edu/abs/2011MNRAS.415...11D} {415, 11}

\bibitem[\protect\citeauthoryear{{Dolag} \& {Stasyszyn}}{{Dolag} \&
  {Stasyszyn}}{2009}]{DolagSta2009}
{Dolag} K.,  {Stasyszyn} F.,  2009, \mn@doi [\mnras]
  {10.1111/j.1365-2966.2009.15181.x}, \href
  {http://adsabs.harvard.edu/abs/2009MNRAS.398.1678D} {398, 1678}

\bibitem[\protect\citeauthoryear{{Dolag}, {Jubelgas}, {Springel}, {Borgani}  \&
  {Rasia}}{{Dolag} et~al.}{2004}]{Dolag2004}
{Dolag} K.,  {Jubelgas} M.,  {Springel} V.,  {Borgani} S.,   {Rasia} E.,  2004,
  \mn@doi [\apjl] {10.1086/420966}, \href
  {http://adsabs.harvard.edu/abs/2004ApJ...606L..97D} {606, L97}

\bibitem[\protect\citeauthoryear{{Dolag}, {Grasso}, {Springel}  \&
  {Tkachev}}{{Dolag} et~al.}{2005}]{dolag2005}
{Dolag} K.,  {Grasso} D.,  {Springel} V.,   {Tkachev} I.,  2005, \mn@doi
  [\jcap] {10.1088/1475-7516/2005/01/009}, \href
  {http://cdsads.u-strasbg.fr/cgi-bin/nph-bib_query?bibcode=2005JCAP...01..009D&db_key=AST}
  {1, 9}

\bibitem[\protect\citeauthoryear{{Dole} et~al.,}{{Dole}
  et~al.}{2006}]{Dole2006}
{Dole} H.,  et~al., 2006, \mn@doi [\aap] {10.1051/0004-6361:20054446}, \href
  {http://adsabs.harvard.edu/abs/2006A%26A...451..417D} {451, 417}

\bibitem[\protect\citeauthoryear{{Draine} \& {Li}}{{Draine} \&
  {Li}}{2007}]{Draine2007}
{Draine} B.~T.,  {Li} A.,  2007, \mn@doi [\apj] {10.1086/511055}, \href
  {http://adsabs.harvard.edu/abs/2007ApJ...657..810D} {657, 810}

\bibitem[\protect\citeauthoryear{{Elbaz}, {Hwang}, {Magnelli}, {Daddi}
  et~al.}{{Elbaz} et~al.}{2010}]{Elbaz10}
{Elbaz} D.,  {Hwang} H.~S.,  {Magnelli} B.,  {Daddi} E.,   et~al., 2010,
  \mn@doi [\aap] {10.1051/0004-6361/201014687}, \href
  {http://adsabs.harvard.edu/abs/2010A%26A...518L..29E} {518, L29}

\bibitem[\protect\citeauthoryear{{Fabjan}, {Borgani}, {Tornatore}, {Saro},
  {Murante}  \& {Dolag}}{{Fabjan} et~al.}{2010}]{Fabjan}
{Fabjan} D.,  {Borgani} S.,  {Tornatore} L.,  {Saro} A.,  {Murante} G.,
  {Dolag} K.,  2010, \mn@doi [\mnras] {10.1111/j.1365-2966.2009.15794.x}, \href
  {http://adsabs.harvard.edu/abs/2010MNRAS.401.1670F} {401, 1670}

\bibitem[\protect\citeauthoryear{{Fontanot}, {Cristiani}, {Santini}, {Fontana},
  {Grazian}  \& {Somerville}}{{Fontanot} et~al.}{2012}]{Fontanot2012}
{Fontanot} F.,  {Cristiani} S.,  {Santini} P.,  {Fontana} A.,  {Grazian} A.,
  {Somerville} R.~S.,  2012, \mn@doi [\mnras]
  {10.1111/j.1365-2966.2011.20294.x}, \href
  {http://adsabs.harvard.edu/abs/2012MNRAS.421..241F} {421, 241}

\bibitem[\protect\citeauthoryear{{Gallego}, {Zamorano}, {Aragon-Salamanca}  \&
  {Rego}}{{Gallego} et~al.}{1995}]{Gallego1995}
{Gallego} J.,  {Zamorano} J.,  {Aragon-Salamanca} A.,   {Rego} M.,  1995,
  \mn@doi [\apjl] {10.1086/309804}, \href
  {http://adsabs.harvard.edu/abs/1995ApJ...455L...1G} {455, L1}

\bibitem[\protect\citeauthoryear{{Glazebrook}, {Blake}, {Economou}, {Lilly}  \&
  {Colless}}{{Glazebrook} et~al.}{1999}]{Glazebrook1999}
{Glazebrook} K.,  {Blake} C.,  {Economou} F.,  {Lilly} S.,   {Colless} M.,
  1999, \mn@doi [\mnras] {10.1046/j.1365-8711.1999.02576.x}, \href
  {http://adsabs.harvard.edu/abs/1999MNRAS.306..843G} {306, 843}

\bibitem[\protect\citeauthoryear{{Gruppioni}, {Pozzi}, {Rodighiero},
  {Delvecchio}  et~al.}{{Gruppioni} et~al.}{2013}]{Gruppionis13}
{Gruppioni} C.,  {Pozzi} F.,  {Rodighiero} G.,  {Delvecchio} I.,   et~al.,
  2013, \mn@doi [\mnras] {10.1093/mnras/stt308}, \href
  {http://adsabs.harvard.edu/abs/2013MNRAS.432...23G} {432, 23}

\bibitem[\protect\citeauthoryear{{Hanish} et~al.,}{{Hanish}
  et~al.}{2006}]{Hanish2006}
{Hanish} D.~J.,  et~al., 2006, \mn@doi [\apj] {10.1086/504681}, \href
  {http://adsabs.harvard.edu/abs/2006ApJ...649..150H} {649, 150}

\bibitem[\protect\citeauthoryear{{Hao}, {Kennicutt}, {Johnson}, {Calzetti},
  {Dale}  \& {Moustakas}}{{Hao} et~al.}{2011}]{HaoKen}
{Hao} C.-N.,  {Kennicutt} R.~C.,  {Johnson} B.~D.,  {Calzetti} D.,  {Dale}
  D.~A.,   {Moustakas} J.,  2011, \mn@doi [\apj] {10.1088/0004-637X/741/2/124},
  \href {http://adsabs.harvard.edu/abs/2011ApJ...741..124H} {741, 124}

\bibitem[\protect\citeauthoryear{{Hayes}, {Schaerer}  \& {{\"O}stlin}}{{Hayes}
  et~al.}{2010}]{Hayes2010}
{Hayes} M.,  {Schaerer} D.,   {{\"O}stlin} G.,  2010, \mn@doi [\aap]
  {10.1051/0004-6361/200913217}, \href
  {http://adsabs.harvard.edu/abs/2010A%26A...509L...5H} {509, L5}

\bibitem[\protect\citeauthoryear{{Helou}}{{Helou}}{1986}]{Helou1986}
{Helou} G.,  1986, \mn@doi [\apjl] {10.1086/184793}, \href
  {http://adsabs.harvard.edu/abs/1986ApJ...311L..33H} {311, L33}

\bibitem[\protect\citeauthoryear{{Hirashita}, {Buat}  \& {Inoue}}{{Hirashita}
  et~al.}{2003}]{Hira03}
{Hirashita} H.,  {Buat} V.,   {Inoue} A.~K.,  2003, \mn@doi [\aap]
  {10.1051/0004-6361:20031144}, \href
  {http://adsabs.harvard.edu/abs/2003A%26A...410...83H} {410, 83}

\bibitem[\protect\citeauthoryear{{Hopkins}, {Connolly}  \& {Szalay}}{{Hopkins}
  et~al.}{2000}]{Hopkins2000}
{Hopkins} A.~M.,  {Connolly} A.~J.,   {Szalay} A.~S.,  2000, \mn@doi [\aj]
  {10.1086/316857}, \href {http://adsabs.harvard.edu/abs/2000AJ....120.2843H}
  {120, 2843}

\bibitem[\protect\citeauthoryear{{Hopkins}, {Connolly}, {Haarsma}  \&
  {Cram}}{{Hopkins} et~al.}{2001}]{Hopkins01}
{Hopkins} A.~M.,  {Connolly} A.~J.,  {Haarsma} D.~B.,   {Cram} L.~E.,  2001,
  \mn@doi [\aj] {10.1086/321113}, \href
  {http://adsabs.harvard.edu/abs/2001AJ....122..288H} {122, 288}

\bibitem[\protect\citeauthoryear{{Iannuzzi} \& {Dolag}}{{Iannuzzi} \&
  {Dolag}}{2011}]{iannuzzi11}
{Iannuzzi} F.,  {Dolag} K.,  2011, \mn@doi [\mnras]
  {10.1111/j.1365-2966.2011.19446.x}, \href
  {http://adsabs.harvard.edu/abs/2011MNRAS.417.2846I} {417, 2846}

\bibitem[\protect\citeauthoryear{{Katsianis}, {Tescari}  \&
  {Wyithe}}{{Katsianis} et~al.}{2015}]{Katsianis2014}
{Katsianis} A.,  {Tescari} E.,   {Wyithe} J.~S.~B.,  2015, \mn@doi [\mnras]
  {10.1093/mnras/stv160}, \href
  {http://adsabs.harvard.edu/abs/2015MNRAS.448.3001K} {448, 3001}

\bibitem[\protect\citeauthoryear{{Katsianis}, {Tescari}  \&
  {Wyithe}}{{Katsianis} et~al.}{2016}]{Katsianis2015}
{Katsianis} A.,  {Tescari} E.,   {Wyithe} J.~S.~B.,  2016, \mn@doi [\pasa]
  {10.1017/pasa.2016.21}, \href
  {http://adsabs.harvard.edu/abs/2016PASA...33...29K} {33, e029}

\bibitem[\protect\citeauthoryear{{Kennicutt}}{{Kennicutt}}{1983}]{Ken83}
{Kennicutt} Jr. R.~C.,  1983, \mn@doi [\apj] {10.1086/161261}, \href
  {http://adsabs.harvard.edu/abs/1983ApJ...272...54K} {272, 54}

\bibitem[\protect\citeauthoryear{{Kennicutt}}{{Kennicutt}}{1998a}]{kennicutt1998}
{Kennicutt} Jr. R.~C.,  1998a, \mn@doi [\araa]
  {10.1146/annurev.astro.36.1.189}, \href
  {http://adsabs.harvard.edu/abs/1998ARA%26A..36..189K} {36, 189}

\bibitem[\protect\citeauthoryear{{Kennicutt}}{{Kennicutt}}{1998b}]{Ken98a}
{Kennicutt} Jr. R.~C.,  1998b, \mn@doi [\apj] {10.1086/305588}, \href
  {http://adsabs.harvard.edu/abs/1998ApJ...498..541K} {498, 541}

\bibitem[\protect\citeauthoryear{{Kewley}, {Geller}  \& {Jansen}}{{Kewley}
  et~al.}{2004}]{Kewley2004}
{Kewley} L.~J.,  {Geller} M.~J.,   {Jansen} R.~A.,  2004, \mn@doi [\aj]
  {10.1086/382723}, \href {http://adsabs.harvard.edu/abs/2004AJ....127.2002K}
  {127, 2002}

\bibitem[\protect\citeauthoryear{{Le Floc'h}, {Papovich}, {Dole}, {Bell}
  et~al.}{{Le Floc'h} et~al.}{2005}]{LeFloch}
{Le Floc'h} E.,  {Papovich} C.,  {Dole} H.,  {Bell} E.~F.,   et~al., 2005,
  \mn@doi [\apj] {10.1086/432789}, \href
  {http://adsabs.harvard.edu/abs/2005ApJ...632..169L} {632, 169}

\bibitem[\protect\citeauthoryear{{Ly}, {Lee}, {Dale}, {Momcheva}, {Salim},
  {Staudaher}, {Moore}  \& {Finn}}{{Ly} et~al.}{2011}]{Ly11}
{Ly} C.,  {Lee} J.~C.,  {Dale} D.~A.,  {Momcheva} I.,  {Salim} S.,  {Staudaher}
  S.,  {Moore} C.~A.,   {Finn} R.,  2011, \mn@doi [\apj]
  {10.1088/0004-637X/726/2/109}, \href
  {http://adsabs.harvard.edu/abs/2011ApJ...726..109L} {726, 109}

\bibitem[\protect\citeauthoryear{{Madau} \& {Dickinson}}{{Madau} \&
  {Dickinson}}{2014}]{Madau2014}
{Madau} P.,  {Dickinson} M.,  2014, \mn@doi [\araa]
  {10.1146/annurev-astro-081811-125615}, \href
  {http://adsabs.harvard.edu/abs/2014ARA%26A..52..415M} {52, 415}

\bibitem[\protect\citeauthoryear{{Madau}, {Pozzetti}  \& {Dickinson}}{{Madau}
  et~al.}{1998}]{madau1998}
{Madau} P.,  {Pozzetti} L.,   {Dickinson} M.,  1998, \mn@doi [\apj]
  {10.1086/305523}, \href {http://adsabs.harvard.edu/abs/1998ApJ...498..106M}
  {498, 106}

\bibitem[\protect\citeauthoryear{{Magnelli}, {Elbaz}, {Chary}, {Dickinson}, {Le
  Borgne}, {Frayer}  \& {Willmer}}{{Magnelli} et~al.}{2011}]{Magenlli11}
{Magnelli} B.,  {Elbaz} D.,  {Chary} R.~R.,  {Dickinson} M.,  {Le Borgne} D.,
  {Frayer} D.~T.,   {Willmer} C.~N.~A.,  2011, \mn@doi [\aap]
  {10.1051/0004-6361/200913941}, \href
  {http://adsabs.harvard.edu/abs/2011A%26A...528A..35M} {528, A35}

\bibitem[\protect\citeauthoryear{{Magnelli}, {Popesso}, {Berta}, {Pozzi}
  et~al.}{{Magnelli} et~al.}{2013}]{Magnelli2013}
{Magnelli} B.,  {Popesso} P.,  {Berta} S.,  {Pozzi} F.,   et~al., 2013, \mn@doi
  [\aap] {10.1051/0004-6361/201321371}, \href
  {http://adsabs.harvard.edu/abs/2013A%26A...553A.132M} {553, A132}

\bibitem[\protect\citeauthoryear{{Maio} \& {Tescari}}{{Maio} \&
  {Tescari}}{2015}]{MaioTe2015}
{Maio} U.,  {Tescari} E.,  2015, \mn@doi [\mnras] {10.1093/mnras/stv1714},
  \href {http://adsabs.harvard.edu/abs/2015MNRAS.453.3798M} {453, 3798}

\bibitem[\protect\citeauthoryear{{Maio}, {Dolag}, {Ciardi}  \&
  {Tornatore}}{{Maio} et~al.}{2007}]{maio2007}
{Maio} U.,  {Dolag} K.,  {Ciardi} B.,   {Tornatore} L.,  2007, \mn@doi [\mnras]
  {10.1111/j.1365-2966.2007.12016.x}, \href
  {http://adsabs.harvard.edu/abs/2007MNRAS.379..963M} {379, 963}

\bibitem[\protect\citeauthoryear{{Martin}}{{Martin}}{2005}]{Martin05}
{Martin} C.~L.,  2005, \mn@doi [\apj] {10.1086/427277}, \href
  {http://adsabs.harvard.edu/abs/2005ApJ...621..227M} {621, 227}

\bibitem[\protect\citeauthoryear{{Meurer}, {Heckman}  \& {Calzetti}}{{Meurer}
  et~al.}{1999}]{meurer1999}
{Meurer} G.~R.,  {Heckman} T.~M.,   {Calzetti} D.,  1999, \mn@doi [\apj]
  {10.1086/307523}, \href {http://adsabs.harvard.edu/abs/1999ApJ...521...64M}
  {521, 64}

\bibitem[\protect\citeauthoryear{{Moorwood}, {van der Werf}, {Cuby}  \&
  {Oliva}}{{Moorwood} et~al.}{2000}]{Moorwood2000}
{Moorwood} A.~F.~M.,  {van der Werf} P.~P.,  {Cuby} J.~G.,   {Oliva} E.,  2000,
  \aap, \href {http://adsabs.harvard.edu/abs/2000A%26A...362....9M} {362, 9}

\bibitem[\protect\citeauthoryear{{Moustakas}, {Kennicutt}  \&
  {Tremonti}}{{Moustakas} et~al.}{2006}]{Moustakas2006}
{Moustakas} J.,  {Kennicutt} Jr. R.~C.,   {Tremonti} C.~A.,  2006, \mn@doi
  [\apj] {10.1086/500964}, \href
  {http://adsabs.harvard.edu/abs/2006ApJ...642..775M} {642, 775}

\bibitem[\protect\citeauthoryear{{Oesch} et~al.,}{{Oesch}
  et~al.}{2010}]{Oesch2010}
{Oesch} P.~A.,  et~al., 2010, \mn@doi [\apjl] {10.1088/2041-8205/725/2/L150},
  \href {http://adsabs.harvard.edu/abs/2010ApJ...725L.150O} {725, L150}

\bibitem[\protect\citeauthoryear{{Parsa}, {Dunlop}, {McLure}  \&
  {Mortlock}}{{Parsa} et~al.}{2015}]{Parsa2015}
{Parsa} S.,  {Dunlop} J.~S.,  {McLure} R.~J.,   {Mortlock} A.,  2015, preprint,
  \href {http://adsabs.harvard.edu/abs/2015arXiv150705629P} {} (\mn@eprint
  {arXiv} {1507.05629})

\bibitem[\protect\citeauthoryear{{P{\'e}rez-Gonz{\'a}lez}, {Zamorano},
  {Gallego}, {Arag{\'o}n-Salamanca}  \& {Gil de Paz}}{{P{\'e}rez-Gonz{\'a}lez}
  et~al.}{2003}]{Pg2003}
{P{\'e}rez-Gonz{\'a}lez} P.~G.,  {Zamorano} J.,  {Gallego} J.,
  {Arag{\'o}n-Salamanca} A.,   {Gil de Paz} A.,  2003, \mn@doi [\apj]
  {10.1086/375364}, \href {http://adsabs.harvard.edu/abs/2003ApJ...591..827P}
  {591, 827}

\bibitem[\protect\citeauthoryear{{Pettini}, {Kellogg}, {Steidel}, {Dickinson},
  {Adelberger}  \& {Giavalisco}}{{Pettini} et~al.}{1998}]{Pettini1998}
{Pettini} M.,  {Kellogg} M.,  {Steidel} C.~C.,  {Dickinson} M.,  {Adelberger}
  K.~L.,   {Giavalisco} M.,  1998, \mn@doi [\apj] {10.1086/306431}, \href
  {http://adsabs.harvard.edu/abs/1998ApJ...508..539P} {508, 539}

\bibitem[\protect\citeauthoryear{{Planelles}, {Borgani}, {Dolag}, {Ettori},
  {Fabjan}, {Murante}  \& {Tornatore}}{{Planelles} et~al.}{2012}]{susana13}
{Planelles} S.,  {Borgani} S.,  {Dolag} K.,  {Ettori} S.,  {Fabjan} D.,
  {Murante} G.,   {Tornatore} L.,  2012, preprint, \href
  {http://adsabs.harvard.edu/abs/2012arXiv1209.5058P} {} (\mn@eprint {arXiv}
  {1209.5058})

\bibitem[\protect\citeauthoryear{{Puchwein} \& {Springel}}{{Puchwein} \&
  {Springel}}{2013}]{PuchweinSpri12}
{Puchwein} E.,  {Springel} V.,  2013, \mn@doi [\mnras] {10.1093/mnras/sts243},
  \href {http://adsabs.harvard.edu/abs/2013MNRAS.428.2966P} {428, 2966}

\bibitem[\protect\citeauthoryear{{Reddy} \& {Steidel}}{{Reddy} \&
  {Steidel}}{2009}]{Reddy2009}
{Reddy} N.~A.,  {Steidel} C.~C.,  2009, \mn@doi [\apj]
  {10.1088/0004-637X/692/1/778}, \href
  {http://adsabs.harvard.edu/abs/2009ApJ...692..778R} {692, 778}

\bibitem[\protect\citeauthoryear{{Reddy}, {Steidel}, {Pettini}, {Adelberger},
  {Shapley}, {Erb}  \& {Dickinson}}{{Reddy} et~al.}{2008}]{Reddy2008}
{Reddy} N.~A.,  {Steidel} C.~C.,  {Pettini} M.,  {Adelberger} K.~L.,  {Shapley}
  A.~E.,  {Erb} D.~K.,   {Dickinson} M.,  2008, \mn@doi [\apjs]
  {10.1086/521105}, \href {http://adsabs.harvard.edu/abs/2008ApJS..175...48R}
  {175, 48}

\bibitem[\protect\citeauthoryear{{Rex}, {Rawle}, {Egami},
  {P{\'e}rez-Gonz{\'a}lez}  et~al.}{{Rex} et~al.}{2010}]{Rex10}
{Rex} M.,  {Rawle} T.~D.,  {Egami} E.,  {P{\'e}rez-Gonz{\'a}lez} P.~G.,
  et~al., 2010, \mn@doi [\aap] {10.1051/0004-6361/201014693}, \href
  {http://adsabs.harvard.edu/abs/2010A%26A...518L..13R} {518, L13}

\bibitem[\protect\citeauthoryear{{Rujopakarn}, {Rieke}, {Eisenstein}  \&
  {Juneau}}{{Rujopakarn} et~al.}{2011}]{Ru2011}
{Rujopakarn} W.,  {Rieke} G.~H.,  {Eisenstein} D.~J.,   {Juneau} S.,  2011,
  \mn@doi [\apj] {10.1088/0004-637X/726/2/93}, \href
  {http://adsabs.harvard.edu/abs/2011ApJ...726...93R} {726, 93}

\bibitem[\protect\citeauthoryear{{Salpeter}}{{Salpeter}}{1955}]{salpeter55}
{Salpeter} E.~E.,  1955, \apj, \href
  {http://adsabs.harvard.edu/abs/1955ApJ...121..161S} {121, 161}

\bibitem[\protect\citeauthoryear{{Schechter}}{{Schechter}}{1976}]{schechter1976}
{Schechter} P.,  1976, \mn@doi [\apj] {10.1086/154079}, \href
  {http://adsabs.harvard.edu/abs/1976ApJ...203..297S} {203, 297}

\bibitem[\protect\citeauthoryear{{Smit}, {Bouwens}, {Franx}, {Illingworth},
  {Labb{\'e}}, {Oesch}  \& {van Dokkum}}{{Smit} et~al.}{2012}]{smit12}
{Smit} R.,  {Bouwens} R.~J.,  {Franx} M.,  {Illingworth} G.~D.,  {Labb{\'e}}
  I.,  {Oesch} P.~A.,   {van Dokkum} P.~G.,  2012, \mn@doi [\apj]
  {10.1088/0004-637X/756/1/14}, \href
  {http://adsabs.harvard.edu/abs/2012ApJ...756...14S} {756, 14}

\bibitem[\protect\citeauthoryear{{Sobral}, {Smail}, {Best}, {Geach}, {Matsuda},
  {Stott}, {Cirasuolo}  \& {Kurk}}{{Sobral} et~al.}{2013}]{Sobral2013}
{Sobral} D.,  {Smail} I.,  {Best} P.~N.,  {Geach} J.~E.,  {Matsuda} Y.,
  {Stott} J.~P.,  {Cirasuolo} M.,   {Kurk} J.,  2013, \mn@doi [\mnras]
  {10.1093/mnras/sts096}, \href
  {http://adsabs.harvard.edu/abs/2013MNRAS.428.1128S} {428, 1128}

\bibitem[\protect\citeauthoryear{{Springel}}{{Springel}}{2005}]{Springel2005}
{Springel} V.,  2005, \mn@doi [\mnras] {10.1111/j.1365-2966.2005.09655.x},
  \href {http://adsabs.harvard.edu/abs/2005MNRAS.364.1105S} {364, 1105}

\bibitem[\protect\citeauthoryear{{Springel} \& {Hernquist}}{{Springel} \&
  {Hernquist}}{2003}]{springel2003}
{Springel} V.,  {Hernquist} L.,  2003, \mn@doi [\mnras]
  {10.1046/j.1365-8711.2003.06206.x}, \href
  {http://adsabs.harvard.edu/abs/2003MNRAS.339..289S} {339, 289}

\bibitem[\protect\citeauthoryear{{Springel} et~al.,}{{Springel}
  et~al.}{2005}]{springeletal05nat}
{Springel} V.,  et~al., 2005, \mn@doi [\nat] {10.1038/nature03597}, \href
  {http://adsabs.harvard.edu/abs/2005Natur.435..629S} {435, 629}

\bibitem[\protect\citeauthoryear{{Sullivan}, {Treyer}, {Ellis}, {Bridges},
  {Milliard}  \& {Donas}}{{Sullivan} et~al.}{2000}]{Sullivan2000}
{Sullivan} M.,  {Treyer} M.~A.,  {Ellis} R.~S.,  {Bridges} T.~J.,  {Milliard}
  B.,   {Donas} J.,  2000, \mn@doi [\mnras] {10.1046/j.1365-8711.2000.03140.x},
  \href {http://adsabs.harvard.edu/abs/2000MNRAS.312..442S} {312, 442}

\bibitem[\protect\citeauthoryear{{Tacchella}, {Trenti}  \&
  {Carollo}}{{Tacchella} et~al.}{2013}]{Tacchella2013}
{Tacchella} S.,  {Trenti} M.,   {Carollo} C.~M.,  2013, \mn@doi [\apjl]
  {10.1088/2041-8205/768/2/L37}, \href
  {http://adsabs.harvard.edu/abs/2013ApJ...768L..37T} {768, L37}

\bibitem[\protect\citeauthoryear{{Teplitz} et~al.,}{{Teplitz}
  et~al.}{2000}]{Teplitz2000}
{Teplitz} H.~I.,  et~al., 2000, \mn@doi [\apj] {10.1086/309539}, \href
  {http://adsabs.harvard.edu/abs/2000ApJ...542...18T} {542, 18}

\bibitem[\protect\citeauthoryear{{Tescari}, {Katsianis}, {Wyithe}, {Dolag},
  {Tornatore}, {Barai}, {Viel}  \& {Borgani}}{{Tescari}
  et~al.}{2014}]{TescariKaW2013}
{Tescari} E.,  {Katsianis} A.,  {Wyithe} J.~S.~B.,  {Dolag} K.,  {Tornatore}
  L.,  {Barai} P.,  {Viel} M.,   {Borgani} S.,  2014, \mn@doi [\mnras]
  {10.1093/mnras/stt2461}, \href
  {http://adsabs.harvard.edu/abs/2014MNRAS.438.3490T} {438, 3490}

\bibitem[\protect\citeauthoryear{{Tornatore}, {Ferrara}  \&
  {Schneider}}{{Tornatore} et~al.}{2007a}]{tornatore07b}
{Tornatore} L.,  {Ferrara} A.,   {Schneider} R.,  2007a, \mn@doi [\mnras]
  {10.1111/j.1365-2966.2007.12215.x}, \href
  {http://adsabs.harvard.edu/abs/2007MNRAS.382..945T} {382, 945}

\bibitem[\protect\citeauthoryear{{Tornatore}, {Borgani}, {Dolag}  \&
  {Matteucci}}{{Tornatore} et~al.}{2007b}]{tornatore07}
{Tornatore} L.,  {Borgani} S.,  {Dolag} K.,   {Matteucci} F.,  2007b, \mn@doi
  [\mnras] {10.1111/j.1365-2966.2007.12070.x}, \href
  {http://adsabs.harvard.edu/abs/2007MNRAS.382.1050T} {382, 1050}

\bibitem[\protect\citeauthoryear{{Tresse}, {Maddox}, {Le F{\`e}vre}  \&
  {Cuby}}{{Tresse} et~al.}{2002}]{Tresse2002}
{Tresse} L.,  {Maddox} S.~J.,  {Le F{\`e}vre} O.,   {Cuby} J.-G.,  2002,
  \mn@doi [\mnras] {10.1046/j.1365-8711.2002.05919.x}, \href
  {http://adsabs.harvard.edu/abs/2002MNRAS.337..369T} {337, 369}

\bibitem[\protect\citeauthoryear{{Wiersma}, {Schaye}  \& {Smith}}{{Wiersma}
  et~al.}{2009}]{wiersma09}
{Wiersma} R.~P.~C.,  {Schaye} J.,   {Smith} B.~D.,  2009, \mn@doi [\mnras]
  {10.1111/j.1365-2966.2008.14191.x}, \href
  {http://adsabs.harvard.edu/abs/2009MNRAS.393...99W} {393, 99}

\bibitem[\protect\citeauthoryear{{Yan}, {Windhorst}  \& {Cohen}}{{Yan}
  et~al.}{2003}]{Yan2003}
{Yan} H.,  {Windhorst} R.~A.,   {Cohen} S.~H.,  2003, \mn@doi [\apjl]
  {10.1086/374371}, \href {http://adsabs.harvard.edu/abs/2003ApJ...585L..93Y}
  {585, L93}

\bibitem[\protect\citeauthoryear{{van der Burg}, {Hildebrandt}  \&
  {Erben}}{{van der Burg} et~al.}{2010}]{VAN2010}
{van der Burg} R.~F.~J.,  {Hildebrandt} H.,   {Erben} T.,  2010, \mn@doi [\aap]
  {10.1051/0004-6361/200913812}, \href
  {http://adsabs.harvard.edu/abs/2010A%26A...523A..74V} {523, A74}

\makeatother
\end{thebibliography}

\label{lastpage}
\end{document}